%% file: main.tex
\documentclass[journal=jpcafh]{achemso}
\setkeys{acs}{articletitle = true}
\SectionNumbersOn
\usepackage{amsmath}
\usepackage[nolist]{acronym}
\usepackage{chemformula}
\usepackage{braket}

\usepackage{xcolor}
\newcommand \ZW[1]{{\color{black}#1}}

\usepackage{enumitem}

\usepackage{caption}
\usepackage{subcaption}
\usepackage{graphicx}
\usepackage{wrapfig}

\makeatletter
\newcounter{subsubsubsection}[subsubsection]

\newcommand\subsubsubsection{\@startsection{subsubsubsection}{4}{\parindent}%
                                    {3.25ex \@plus1ex \@minus .2ex}%
                                    {-1em}%
                                    {\normalfont\normalsize\bfseries}}
\newcommand*\l@subsubsubsection{\@dottedtocline{4}{12em}{5em}}
\newcommand*{\subsubsubsectionmark}[1]{}
\makeatother

\title{Real-Time Coupled Cluster Theory with Approximate Triples}
\author{Zhe Wang}
\affiliation{Department of Chemistry, Virginia Tech, Blacksburg, VA 24061, USA}
\author{H{\aa}kon Emil Kristiansen}
\affiliation{ Hylleraas Centre for Quantum Molecular Sciences, Department of Chemistry, University of Oslo, P.O. Box 1033 Blindern, N-0315 Oslo, Norway}
\author{Thomas Bondo Pedersen}
\affiliation{ Hylleraas Centre for Quantum Molecular Sciences, Department of Chemistry, University of Oslo, P.O. Box 1033 Blindern, N-0315 Oslo, Norway}
\author{T. Daniel Crawford}
\affiliation{Department of Chemistry, Virginia Tech, Blacksburg, VA 24061, USA}
\email{crawdad@vt.edu}
\begin{document}
\input{abstract.tex}
\input{intro.tex}
\input{theory.tex}
\input{comp.tex}

\input{results.tex}
\input{conc.tex}
\input{extra.tex}
%TBP:
\appendix
\input{appendix.tex}

\bibliography{refs.bib}
%\newpage
%\input{toc}
\end{document}

%% file: abstract.tex
\section*{Abstract}

In order to explore the effects of high levels of electron correlation on the real-time coupled cluster
formalism and algorithmic behavior, we introduce a time-dependent implementation of the CC3 singles, doubles
and approximate triples method.  We demonstrate the validity of our derivation and implementation using
specific applications of frequency-dependent properties.  Terms with triples are calculated and added to the
existing CCSD equations, giving the method a nominal ${\cal O}(N^{7})$ scaling.  We also use a graphics
processing unit (GPU) accelerated implementation to reduce the computational cost, which we find can speed up
the calculation by up to a factor of 17 for test cases of water clusters.  In addition, we compare the impact
of using single-precision arithmetic compared to conventional double-precision arithmetic.  We find no
significant difference in polarizabilities and optical-rotation tensor results, but a somewhat larger error
for first hyperpolarizabilities.  Compared to linear response (LR) CC3 results, the percentage errors of
RT-CC3 polarizabilities and RT-CC3 first hyperpolarizabilities are under 0.1\% and 1\%, respectively, for a
water-molecule test case in a double-zeta basis set.  Furthermore, we compare the dynamic polarizabilities
obtained using RT-CC3, RT-CCSD, and time-dependent nonorthogonal orbital-optimized coupled cluster doubles
(TDNOCCD), in order to examine the performance of RT-CC3 and the orbital-optimization effect using a set of
ten-electron systems.

%% file: intro.tex
\section{Introduction} \label{ch4-intro}

Coupled cluster (CC) theory\cite{Gauss98,Bartlett07,Bartlett10,Crawford2000,Shavitt2009} has been proven to be one of the
most accurate and robust methods for treating electron correlation effects for a wide range of molecular properties, from
ground-state energies and molecular structure to thermodynamic properties, electronic and vibrational spectra, response
properties, and more.  While CC methods are exact in the untruncated (Born-Openheimer, non-relativistic) limit,
restriction of the wave function to modest excitation levels is ultimately necessary for practical
applications.\cite{Gauss00:properties}  While the inclusion of single and double excitations (CCSD) has been widely used
and affirmed to be effective and efficient (with ${\cal O}(N^6)$ scaling, where $N$ is related to the size of the system),
higher levels are often required to achieve the accuracy needed for quantitative comparison to experiment.  However,
extension of the wave function even to include just full triples (CCSDT) is impractical for most applications due to the
${\cal O}(N^8)$ scaling of the method.\cite{Hoffmann86,Noga1987}

Over the last several decades, researchers have explored a range of approximations to the full CCSDT
model,\cite{Lee1984,Urban1985,Noga1987ccsdtn,Raghavachari1989,Stanton1997,Crawford98:aT,Kucharski98,Kowalski2000,Piecuch2002} the most successful
of which is the CCSD(T) approach.\cite{Raghavachari1989} 
In this approximation, the converged CCSD singles and doubles amplitudes are used to
estimate the triples in a non-iterative manner by adding dominant terms in the many-body perturbation theory (MBPT)
expansion of the correlation energy. In particular, for Hartree-Fock reference determinants, the (T) correction includes
triples contributions to the energy involving the doubles at fourth order and the singles at fifth order (which becomes
fourth order for non-Brillouin references).  The leads to a non-iterative ${\cal O}(N^7)$ method which is commonly
referred to as the ``gold standard'' of quantum chemistry.

For response properties, such as dynamic polarizabilities, which require a time-dependent formulation, the CCSD(T) model
suffers from the lack of coupling between the triples and the lower-excitation amplitudes.  Thus, the triples do not
respond directly to an external field, for example, yielding the same pole structure of the CCSD response functions.  This
was the motivation for the development of the CC3 approach by Koch, Christiansen, J{\o}rgensen, and
co-workers.\cite{Christiansen1995CC3,Koch1997} CC3 is an iterative model that treats singles uniquely as zeroth-order
parameters to approximate orbital relaxation effects, with triples being correct to the second order of MBPT. In each
iteration, the approximate triples are calculated and used to determine their contributions to the singles and doubles, in
order to correct the energy to fifth order.  In this manner, contractions with a scaling of ${\cal O}(N^{7})$ occur involving the
triples amplitudes, but the complete triples tensor need not to be stored at any time during the calculation for a time-independent
formulation.

CC3 can provide comparable results compared to other iterative and non-iterative approximate triples models for
time-independent properties.\cite{Koch1997}  More crucially, CC3-level time-/frequency-dependent properties, including
response functions, can be derived.\cite{Koch1997, Christiansen1995CC3, Christiansen1998triple, Gauss1998} If considered only in comparison with other
iterative models, counting singles as zero-order parameters underscores the greater importance of singles for properties
related to the perturbation of an external electromagnetic field and excited states. This makes it an exceptional
candidate to be combined with, for example, linear- and high-order response
theory,\cite{Olsen1985,Sekino1984,Helgaker12,Norman18} and real-time (RT)
methods,\cite{Goings2018,Li2020,Ofstad2023_review,Huber2011,Kvaal2012,Sato2018,Pedersen2019,Nascimento2016,Nascimento2019,Pedersen2021,Wang2022,Peyton2023} in order to
include higher excitations beyond CCSD.  Implementations of CC3 response functions have been reported in comparison to
other iterative triples models.\cite{Christiansen1995CC3} For example, excitation energies of small molecules are
significantly improved by CC3 relative to CCSD, and were found to be comparable to full CCSDT at greatly reduced
cost.\cite{Hald2002} Second-order properties, such as polarizabilities, were also obtained from CC3 linear response
functions and yielded good agreement with experimental data for the systems considered.\cite{Hald2003}

Here, we report the first implementation of the RT-CC3 method, which is built upon our existing RT-CC
framework.\cite{Wang2022,Peyton2023} A similar RT-CC method with explicitly time-dependent orbitals and 
perturbation-based treatment of triple excitation amplitudes was presented recently by \citeauthor{Pathak2021}\cite{Pathak2021,Pathak2022}
We have computed RT-CC3 absorption spectra for several small molecular test cases for
comparison to excitation energies and dipole strengths obtained from conventional time-independent response theory to
validate the RT-CC3 implementation.  For higher-order properties, instead of using the derivatives of the time-averaged
quasienergy, \textit{i.e.}, a response formulation, we have used time-dependent finite-difference
methods\cite{Perrone1975} to obtain polarizabilities, the $G'$ tensor related to optical rotation, first
hyperpolarizabilities, and the quadratic response function 
$\langle\!\langle \boldsymbol{\hat{m}};\boldsymbol{\hat{\mu}},\boldsymbol{\hat{\mu}}\rangle\!\rangle_{\omega,\omega^\prime}$. 
This approach was first proposed by \citeauthor{Ding2013} in an application of real-time
time-dependent density functional theory (RT-TDDFT), allowing response properties to be obtained from a cohort of
propagations using different (weak) field strengths.~\cite{Ding2013} We find that only relatively short propagations are
required to obtain such properties and that properties related to different orders of response to the same field can be
calculated using the same group of propagations.  We provide details regarding our implementation of the RT-CC3 method,
the accuracy and stability of the corresponding simulations, as well as discussion of the finite-difference methods and
comparison with RT simulations at other levels of theory in subsequent sections.

%% file: theory.tex
\section{Theory} \label{theory_cc3} 
\subsection{Implementation of the RT-CC3 Method}  \label{theory-cc3-1}
The Hamiltonian perturbed by an external field can be defined as
\begin{equation}
\hat{H}  \rightarrow \hat{H} + \beta\hat{V} = \hat{F} + \hat{U} + \beta\hat{V},
\label{eq:cc3-h-pert}
\end{equation}
where $\hat{F}$ is the Fock operator, $\hat{U}$ is the fluctuation potential,
$\beta$ is the field strength, and $\hat{V}$ is the one-electron 
perturbation operator, such as an electromagnetic field, with strength, $\beta$.
In RT-CC methods, the
differential equations of the time-dependent $\hat{T}$ and $\hat{\Lambda}$
amplitudes can be derived from the time-dependent Schr\"odinger equation by
explicitly differentiating the amplitudes with respect to time, \textit{viz.},
\begin{equation}
i\dot{t}_{\mu} = \bra{\mu}\bar{H}\ket{\Phi_{0}}
\label{eq:cc3-rt-t}
\end{equation}
and
\begin{equation}
-i\dot{\lambda}_{\mu} = \bra{\Phi_{0}}(1+\hat{\Lambda})[\bar{H}, \tau_{\mu}]\ket{\Phi_{0}},
\label{eq:cc3-rt-l}
\end{equation}
where $\bar{H}$ is the similarity-transformed Hamiltonian, $e^{-\hat{T}}\hat{H}e^{\hat{T}}$, 
and $\tau_{\mu}$ is the second-quantized operator that produces the excited determinant $\ket{\mu}$ from the reference $\ket{\Phi_0}$. It is important to note that the
right-hand sides of the two equations are equivalent to the amplitude residuals
in the ground state amplitude equations. Consequently, the computational cost of 
evaluating the right-hand side once is equivalent to the cost of calculating 
the corresponding amplitude residual during a single iteration of the ground-state calculation. 
The amplitudes are inherently complex-valued functions of time.

The complete derivation and the spin-adapted expression of the RT-CC3 equations 
are provided in the Appendix,
while key details of the implementation are highlighted here. For RT-CC3 calculations,
amplitude residuals of singles and doubles can be separated into the CCSD component 
and the contribution from the triples. 
In each time step of the real-time propagation, the CCSD amplitude residuals
are calculated first, followed by the calculation of the contribution from
triples to singles and doubles. Since the $\hat{T}_{3}$ amplitude is a six-index
quantity, storing the entire tensor with a size of $N_{O}^{3}N_{V}^{3}$ is
neither preferable nor feasible due to limited memory, especially when dealing
with large molecular systems and/or large basis sets. The treatment of triples differs
depending on wether the external perturbation is present or absent during the calculation.

Given the general form of CC3 $\hat{T}_{3}$ and $\hat{\Lambda}_{3}$ equations
\begin{equation}
\bra{\mu_{3}}[H, \hat{T}_{2}] + [F + \beta V, \hat{T}_{3}] + \frac{1}{2}[[\beta V, \hat{T}_{2}], \hat{T}_{2}]\ket{\Phi_{0}}=0,
\label{eq:cc3-t3-pert}
\end{equation}
and
\begin{equation}
\bra{\Phi_{0}}\hat{\Lambda}_{1}\hat{H} + \hat{\Lambda}_{2}H + \hat{\Lambda}_{3}(F + \beta V)\ket{\nu_{3}}=0,
\label{eq:cc3-l3-pert}
\end{equation}
respectively, the terms can be rearranged as
\begin{equation}
t_{\mu_{3}}= -\frac{\bra{\mu_{3}}[U, \hat{T}_{2}]\ket{\Phi_{0}}}{\epsilon_{\mu_{3}}}.
\label{eq:cc3-t3-no-pert}
\end{equation}
and
\begin{equation}
\lambda_{\mu_{3}}=-\frac{\bra{\Phi_{0}}\hat{\Lambda}_{1}\hat{H} + \hat{\Lambda}_{2}H\ket{\nu_{3}}}{\epsilon_{\mu_{3}}}
\label{eq:cc3-l3}
\end{equation}
when the perturbation is absent ($\beta=0$). $H$, $F$, $U$, $V$ are the $T_1$-transformed 
Hamiltonian, Fock operator, fluctuation operator, and perturbation operator, respectively, 
with a $T_1$-transformed operator defined as
\begin{equation}
O = e^{-\hat{T}_{1}}\hat{O}e^{\hat{T}_{1}}.
\end{equation}
Any subset of triples can be calculated explicitly with singles and doubles. During each time step,
only a specific subset of triples is calculated on-the-fly when it is needed in a
contraction. For example, in the spin-adapted $\hat{T}_{1}$ equation, 
the contribution of triples can be calculated as
\begin{equation}
\bra{\mu_{1}}[H, \hat{T}_{3}] \ket{\Phi_{0}} = \sum_{jkbc}(t_{ijk}^{abc}-t_{ijk}^{cba})L_{jkbc} \rightarrow t_{i}^{a}.
\label{eq:cc3-x1}
\end{equation}
The subset of triples
corresponding to a certain set of occupied orbitals $i, j, k$ is calculated and
contracted with the subset of integrals corresponding to the same orbitals $j$
and $k$ to calculate its contribution to the $\hat{T}_{1}$ amplitudes with the
occupied orbital $i$. It is important to note that the subset of triples to be
calculated can be a tensor with fixed occupied orbitals $i, j, k$, or fixed
unoccupied orbitals $a, b, c$. The former approach requires performing the
triples calculations and subsequent contractions $N_{O}^{3}$ times, while the
latter requires these calculations $N_{V}^{3}$ times. As $N_{V}$ is typically much
larger than $N_{O}$, it is typically more efficient to compute $N_V^3$ blocks of triples for a given $i, j, k$ combination than to compute
$N_O^3$ blocks for a given $a, b, c$ combination.

When the perturbation is present, the terms $\bra{\mu_{3}}[\beta V, \hat{T}_{3}]\ket{\Phi_{0}}$ 
and $\bra{\mu_{3}}\frac{1}{2}[[\beta V, \hat{T}_{2}], \hat{T}_{2}]\ket{\Phi_{0}}$ in Eq.~(\ref{eq:cc3-t3-pert}),
and $\bra{\Phi_{0}}\hat{\Lambda}_{3} (\beta V) \ket{\nu_{3}}$ in Eq.~(\ref{eq:cc3-l3-pert}) 
must be included. In such cases, the complete set of triples must be calculated for the terms involving
$\hat{T}_{3}$ and $\hat{\Lambda}_{3}$. For the efficency of the calculation, the triples are calculated 
before their first usage in an evaluation of the amplitude residual and are kept in memory or on disk until the end of the 
evaluation. Thus, the calculation requires larger storage space, which could be a 
critical limitation for large molecures and/or basis sets. The calculation of triples and the $V$-dependent
terms will also lead to an increased running time. Similarly, when calculating the one electron density matrix,
the triples are computed once and shared for different blocks of the matrix.
For the real-time propagation,
the $V$-dependent terms do not always need to be calculated throughout the whole propagation. 
If the external field is switched off at a certain time step, the full set of triples and the 
additional terms in the triples equations are no longer needed for the rest of the propagation.

\subsection{Frequency-Dependent Properties from RT Simulations} \label{theory-cc3-2}
\subsubsection{Absorption Spectrum} \label{theory-cc3-21}
For electromagnetic fields in the dipole approximation, the perturbation operator can be specifically written as
\begin{equation}
\hat{V}(t) = -\hat{\mu}\cdot \textbf{E}(t),
\end{equation}
with the system interacting with an external electric field $\textbf{E}(t)$.
Linear absorption spectra can be calculated using the frequency-dependent
counterparts of the time-dependent dipole and electric field, obtained via the
Fourier transform:
\begin{equation}
\tilde{f}(\omega) = \frac{1}{2\pi}\int_{-\infty}^{+\infty}f(t)e^{i\omega t}dt.
\end{equation}
The dipole strength function used here to quantify the probability of the
absorption process is proportional to the imaginary part of the trace of the
dipole polarizability tensor $\boldsymbol{\alpha}(\omega)$,
\begin{equation}
    I(\omega) \propto \mathrm{Im} \left[\sum_{\beta}\alpha_{\beta \beta}(\omega)\right],
\end{equation}
where $\beta$ is the Cartesian axis $x, y, z$. The dipole polarizability $\alpha_{\beta\beta}$ can be
calculated as
\begin{equation}
    \alpha_{\beta\beta}(\omega) = \frac{\tilde{\mu}_{\beta}(\omega)}{\tilde{E}_{\beta}(\omega)}.
\end{equation}

\subsubsection{Dynamic Polarizabilities and Hyperpolarizabilities} \label{theory-cc3-22}
Consider a molecule exposed to a field with the form of
\begin{equation}
E_{\beta}(t) = A_{\beta}\cos(\omega t),
\label{eq:cc3-polar-field}
\end{equation}
where $A_{\beta}$ and $\omega$ are the maximum amplitude and the frequency of
the field, respectively, with $\beta$ being the Cartesian axis that indicates
the direction of the field. Under this electric field, the time-dependent
electric dipole moment can be expanded as (see, e.g., Refs. \citenum{Ding2013} and \citenum{Ofstad2023} for details)
\begin{equation}
\mu_{\alpha}(t) = (\mu_{\alpha})_{0} + \alpha_{\alpha\beta}(\omega)\cos(\omega t)A_{\beta} 
+ \frac{1}{4}[ \beta_{\alpha\beta\beta}(-2\omega;\omega,\omega)\cos(2\omega t) + \beta_{\alpha\beta\beta}(0;\omega,-\omega)]A_{\beta}^{2} + \cdots,
\label{eq:cc3-polar-mu1}
\end{equation}
where $\alpha(\omega)$ is the polarizability, $\beta(-2\omega;\omega,\omega)$
and $\beta(0;\omega,-\omega)$ are the first-hyperpolarizabilities corresponding
to the second-harmonic generation (SHG) and optical rectification (OR),
respectively. Alternatively, if we write the series expansion of the electric
dipole moment as 
\begin{equation}
    \mu_{\alpha}(t) = \mu_{\alpha}^{(0)} + \mu_{\alpha\beta}^{(1)}(t) A_{\beta} + \mu_{\alpha\beta\beta}^{(2)}(t)A_{\beta}^{2} + \cdots,
\label{eq:cc3-polar-mu2}
\end{equation}
and then equate Eqs.~(\ref{eq:cc3-polar-mu1}) and~(\ref{eq:cc3-polar-mu2}), we obtain
\begin{equation}
    \mu_{\alpha\beta}^{(1)}(t) = \alpha_{\alpha\beta}\cos(\omega t),
\label{eq:cc3-polar-alpha}
\end{equation}
and
\begin{equation}
\mu_{\alpha\beta\beta}^{(2)}(t) = \frac{1}{4}[ \beta_{\alpha\beta\beta}(-2\omega;\omega,\omega)\cos(2\omega t) + \beta_{\alpha\beta\beta}(0;\omega,-\omega)].
\label{eq:cc3-polar-beta}
\end{equation}
One way to calculate the first- and second-order dipole moments is using the
(central) finite-difference method, which is commonly employed for numerical
differentiation. To apply it to real-time methods, induced dipole moments from
simulations with different field strengths are required. For instance,
conducting four separate simulations with field strengths of $A$, $-A$, $2A$,
and $-2A$ as the only varying parameter, allows us to express $\mu_{\alpha
\beta}^{(1)}$ and $\mu_{\alpha \beta\beta}^{(2)}$ as
\begin{equation}
\mu_{\alpha \beta}^{(1)}(t)=\frac{8[\mu_{\alpha}(t,A_{\beta})-\mu_{\alpha}(t,-A_{\beta})]-[\mu_{\alpha}(t,2A_{\beta})-\mu_{\alpha}(t,-2A_{\beta})]}{12A_{\beta}},
\end{equation}
and 
\begin{equation}
\mu_{\alpha \beta\beta}^{(2)}(t)=\frac{16[\mu_{\alpha}(t,A_{\beta})+\mu_{\alpha}(t,-A_{\beta})]-[\mu_{\alpha}(t,2A_{\beta})+\mu_{\alpha}(t,-2A_{\beta})]-30\mu^{(0)}_{\alpha}}{24A_{\beta}^2}.
\end{equation}
With the value of $\mu_{\alpha \beta}^{(1)}$ at each time step, we can fit the
trajectory into a cosine curve, as shown in Eq.~(\ref{eq:cc3-polar-alpha}). The
polarizability $\alpha(\omega)$ will be the amplitude of the fitted curve.
Similarly, the trajectory of $\mu_{\alpha \beta\beta}^{(2)}$ can also be fitted
into a curve with the form $1/4[A\cos(\omega t) + B]$, where
$\beta(-2\omega;\omega,\omega)$ and $\beta(0;\omega,-\omega)$ are the values of
$A$ and $B$ respectively. Additional details about the finite difference method
and its application in the real-time framework can be found in
Refs.~\citenum{Ding2013} and \citenum{Ofstad2023}. It is worth noting that although calculating
(hyper-)polarizabilities at each frequency requires four
real-time simulations, each simulation does not need to be as long as the ones
used for calculating the absorption spectrum, where the spectral resolution
inherently depends on the propagation length. Moreover, both polarizabilities and
hyperpolarizabilities at the same frequency can be obtained from the same set
of simulations. The difference lies only in the post-processing steps.

\subsubsection{$G'$ Tensor and Magnetic-/Electric-Dipole Quadratic Response Function} \label{theory-cc3-23}
In addition to the properties associated with the induced electric dipole
moments, the $G'$ tensor that is related to linear chiroptical properties (optical rotation, electronic
circular dichroism, etc.) and the response function 
$\langle\!\langle \hat{m}_\alpha; \hat{\mu}_\beta, \hat{\mu}_\beta \rangle\!\rangle$ are also accessible
under this formalism, in that they are connected to the magnetic dipole moments induced
by external electric fields. Following the same steps as above, we first write the time series expansion 
of the magnetic dipole moment as:
\begin{equation}
m_{\alpha}(t)= m_{\alpha}^{(0)} + m_{\alpha \beta}^{(1)}(t)\ A_{\beta} + m_{\alpha \beta \beta}^{(2)}(t)\ A^{2}_{\beta} + \cdots.
\label{eq:mag-time-exp}
\end{equation}
For the $G'$ tensor corresponding to the first-order induced magnetic dipole moments, we expand $m_\alpha$ as
\begin{equation}
m_{\alpha} (t)= (m_{\alpha} )_{0} +\frac{1}{\omega} G^{'}_{\beta\alpha}(\omega)\dot E_{\beta} + \cdots,
\end{equation}
with the time derivative of the field being $\dot E_{\beta} = -A\omega sin(\omega t)$, 
and then equate it with Eq.~(\ref{eq:mag-time-exp}). $m_{\alpha\beta}^{(1)}$ can thus be written as
\begin{equation}
m_{\alpha \beta}^{(1)}(t)= G^{'}_{\alpha \beta}(\omega)sin(\omega t),
\end{equation}
and calculated as
\begin{equation}
m_{\alpha \beta}^{(1)}(t)=\frac{8[m_{\alpha}(t,A_{\beta})-m_{\alpha}(t,-A_{\beta})]-[m_{\alpha}(t,2A_{\beta})-m_{\alpha}(t,-2A_{\beta})]}{12A_{\beta}}.
\end{equation}
To obtain $\langle\!\langle \hat{m}_{\alpha}; \hat{\mu}_{\beta}, \hat{\mu}_{\beta} \rangle\!\rangle$ 
from the second-order induced magnetic dipole moments, we expand $m_\alpha$ in the 
frequency domain alternatively as
\begin{align}
m_{\alpha}(t) &= m_{\alpha}^{(0)} - \int_{-\infty}^{\infty}\langle\!\langle\hat{m}_{\alpha};\hat{\mu}_{\beta}\rangle\!\rangle_{\omega}
                \tilde{F}(\omega)e^{-i\omega t}d\omega \\ 
		&+ \frac{1}{2}\int\int_{-\infty}^{\infty}\langle\!\langle\hat{m}_{\alpha}; 
		\hat{\mu}_{\beta},\hat{\mu}_{\beta}\rangle\!\rangle_{\omega_{1},\omega_{2}}
		\tilde{F}(\omega_{1})\tilde{F}(\omega_{2})e^{-i(\omega_{1}+\omega_{2})t} d\omega_{1} d\omega_{2} +\cdots.
\label{eq:mag-freq-exp}
\end{align}
By equating Eq.~(\ref{eq:mag-freq-exp}) and Eq.~(\ref{eq:mag-time-exp}), $m_{\alpha \beta \beta}^{(2)}$ can then be written as
\begin{equation}
m_{\alpha \beta \beta}^{(2)}(t) = \frac{1}{4}[\langle\!\langle \hat{m}_\alpha; \hat{\mu}_\beta, \hat{\mu}_\beta \rangle\!\rangle_{\omega, \omega}
                               \cos(2\omega t) + \langle\!\langle \hat{m}_\alpha; \hat{\mu}_\beta, \hat{\mu}_\beta \rangle\!\rangle_{\omega, -\omega}],
\end{equation}
and calculated as
\begin{equation}
m_{\alpha \beta\beta}^{(2)}(t)=\frac{16[m_{\alpha}(t,A_{\beta})+m_{\alpha}(t,-A_{\beta})]-[m_{\alpha}(t,2A_{\beta})+m_{\alpha}(t,-2A_{\beta})]-30m^{(0)}_{\alpha}}{24A_{\beta}^2}.
\end{equation}

For the magnetic dipole moments, no additional modifications to the real-time
framework are needed. Once we calculate the one-electron density, the
electric dipole moment can be obtained by contracting the density with the
electric dipole operator, and the magnetic dipole moment can be obtained by
contracting the density with the magnetic dipole operator. 
It is worth mentioning that the simple contraction of the density with the
magnetic dipole matrix elements is not gauge-invariant since the orbitals are
fixed in our implementation.

\subsubsection{Ramped Continuous Wave} \label{theory-cc3-24}
As shown in Eq.~(\ref{eq:cc3-polar-field}), a cosine wave with a frequency of
$\omega$ is used to calculate the optical properties. In practice, instead of
having the same field from the beginning to the end, a ramped wave is applied,
gradually switching on the field. There are two types of ramped waves that are typically used, 
a linear ramped continuous wave (LRCW)\cite{Ding2013}
\begin{equation}
F_\mathrm{LRCW}=
\left\{
\begin{array}{lcl}
    \frac{t}{t_{r}}\cos(\omega t) & \qquad & 0\le t < t_{r}, \\
    \cos(\omega t) & \qquad & t_{r}\le t \le t_{tot},
\end{array}
\right.
\end{equation}
and a quadratic ramped continuous wave (QRCW)\cite{Ofstad2023}
\begin{equation}
 F_\mathrm{QRCW}=
\left\{
\begin{array}{lcl}
    \frac{2t^2}{t_r^2} \cos(\omega t) & \quad &  0 \leq t < \frac{t_r}{2}, \\
    \left[ 1 -  \frac{2(t-t_r)^2}{t_r^2} \right] \cos(\omega t)  & \quad &  \frac{t_r}{2} \leq t < t_r, \\
    \cos(\omega t)  & \quad &  t_r \leq t \leq t_{tot},
\end{array}
\right.
\end{equation}
where $t_{r}$ is the duration of the ramped field and $t_{tot}$ is the total
length of the simulation. For a field with a frequency $\omega$, an optical
cycle given by
\begin{equation}
t_{c} = \frac{2\pi}{\omega}.
\end{equation}
With $n_r$ and $n_p$ the number of optical cycles used for ramping and for
subsequent propagation at full field strength, the total propagation time
becomes
\begin{equation}
t_{tot} = t_{r} + t_{p} = n_{r}t_{c} + n_{p}t_{c},
\end{equation}
Here, $t_p$ denotes the portion of the propagation utilized for property
fitting.
\citeauthor{Ofstad2023} demonstrated in
Ref.~\citenum{Ofstad2023} that the QRCW can reduce the number of optical cycles
required for both the ramped and subsequent cycles. This reduction is
attributed to the QRCW's more gentle amplification over time in comparison to
the LRCW, resembling an adiabatic switch-on of the field. Moreover, the QRCW is smooth
whereas the LRCW has discontinuous first derivatives at the start and end points of the ramp.
The QRCW thus allows the system to
stabilize more rapidly, even in a shorter time, ensuring that the electrons do not experience
an abrupt perturbation initially. \citeauthor{Ofstad2023} concluded that
for accurate fitting of polarizabilities and first hyperpolarizabilities, one ramped
cycle and one subsequent cycle for curve fitting are sufficient, providing
accurate results compared to linear response theory, which assumes a
monochromatic pulse that is adiabatically switched on by definition. They also demonstrated 
through multiple test cases that the LRCW typically requires at least four subsequent cycles 
following a single ramped cycle to achieve convergence and accuracy. Thus, in 
the present work, we have carried out RT-CC3 calculations with $n_{r}=n_{p}=1$ as the
default values for the QRCW, and $n_{r}=1$, $n_{p}=4$ as the default values for the LRCW in comparison.

%% file: comp.tex
\section{Computational Details} \label{comp_cc3} 
When calculating the absorption spectrum and comparing it with equation-of-motion CC (EOM-CC)
results, an isotropic electric field shaped as a Gaussian function is applied
to the system and shown as
\begin{equation}
\textbf{E}(t) = \mathcal{E}e^{-\frac{(t-\nu)^{2}}{2\sigma^{2}}}\textbf{n},
\end{equation}
where the vector $\textbf{n}$ represents the direction of the field as 
\begin{equation}
\textbf{n} = \frac{1}{\sqrt{3}}(\hat{i}+\hat{j}+\hat{k}).
\end{equation}
The center $\nu$ and the width $\sigma$ of the field are chosen to be 0.01 au and 0.001 au, respectively, to mimic a delta
pulse that is switched on at the beginning of the propagation. For the calculation of RT-CC3/cc-pVDZ absorption spectrum of
\ch{H_{2}O}, the field strength $\mathcal{E}$, step size $h$, and propagation time $t_{f}$ were chosen to be 0.01 au, 0.01
au, and 300 au, respectively.  Pad{\'e} approximants\cite{Bruner2016} were used to improve the resolution of the RT-CC3
spectrum.

To further test the performance of our RT-CC3 implementation, CPU and GPU
calculations were carried out using water monomer, dimer, and trimer systems in
both single- and double-precision. Each CPU calculation was run on a single
node with an AMD EPYC 7702 chip, and each GPU calculation was run on a single
node with an Nvidia Tesla P100 GPU. Tensor manipulation was conducted using
NumPy\cite{Harris2020} and PyTorch\cite{Paszke2019} for the CPU and GPU
calculations, respectively, with similar syntax. Tensor contraction was
performed using {\tt opt\_einsum},\cite{Smith2018} and a PyTorch backend was
specifically employed for the GPU calculation. All calculations kept the $1s$
orbitals of the oxygen atoms frozen.

For calculating dynamic polarizabilities and first hyperpolarizabilities, a set
of RT calculations for the water molecule with the cc-pVDZ basis
set\cite{Dunning1989} were executed using field strengths of $0.002\
\text{au}$, $-0.002\ \text{au}$, $0.004\ \text{au}$, and $-0.004\ \text{au}$ at
both the CC3 and CCSD levels. The step size was set to $0.01\ \text{au}$. The
carrier frequency of the field was set to $0.078\ \text{au}$, which corresponds
to a wavelength of $582\ \text{nm}$ and is lower than the resonance at $0.247\
\text{au}$ for the water molecule. The molecule was subjected to a field in the
$x$, $y$, and $z$ directions individually to obtain the corresponding elements
of the polarizabilities and first hyperpolarizabilities tensors. For the $G'$
tensors and the response function 
$\langle\!\langle\hat{m}_{\alpha};\hat{\mu}_{\beta},\hat{\mu}_{\beta}\rangle\!\rangle$, 
the same electric field was applied to the H$_{2}$ dimer with the
cc-pVDZ basis set. The $G'$ tensor elements and the response functions
were calculated from the induced magnetic dipole moments. The frequency of 
$0.078\ \text{au}$ is below the resonance at $0.367\ \text{au}$. Curve fitting 
was performed using {\tt scipy.optimize.curve\_fit}.\cite{Virtanen2020} 
All calculations were performed on a single Nvidia Tesla P100 GPU, and both 
single- and double-precision calculations were conducted and compared. The results 
from the RT simulations were also compared to reference values obtained 
from the Psi4\cite{Smith2020} and CFOUR\cite{Matthews2020} packages. 

RT-CC methods were also compared to the time-dependent nonorthogonal
orbital-opt\-im\-ized coupled cluster doubles (TDNOCCD) method\cite{Pedersen2001}
for calculating polarizabilities with ten-electron systems including \ch{Ne},
\ch{HF}, \ch{H2O}, \ch{NH3}, and \ch{CH4}. The field strengths of the
propagations were chosen to be $0.001\ \text{au}$, $-0.001\ \text{au}$, $0.002\
\text{au}$, and $-0.002\ \text{au}$. Various frequencies below the resonance of
the corresponding molecule were tested. The basis set was chosen to be
aug-cc-pVDZ\cite{Woon1993} for \ch{HF}, \ch{H2O}, \ch{NH3}, and
\ch{CH4}, and d-aug-cc-pVDZ\cite{Woon1994} for \ch{Ne}. The QRCW method was
utilized for accuracy and efficiency. The length of each propagation is two
optical cycles, depending on the frequency of the field. The time step of the
propagations was $0.01\ \text{au}$. All calculations were performed in
double-precision. 

All calculations were run in PyCC\cite{pycc} with the stationary electric and
magnetic dipole operators extracted from Psi4. The Runge-Kutta fourth-order
integrator\cite{Butcher1996} was used for the RT propagations. For the series
of water clusters, (H$_2$O)$_n$ up to $n=4$, used in
section~\ref{results-cc3-1}, as well as for \ch{H2O} in
section~\ref{results-cc3-22}, the coordinates were provided by \citeauthor{Pokhilko2018}~\cite{Pokhilko2018}
Five ten-electron systems, \ch{Ne}, \ch{HF},
\ch{H2O}, \ch{NH3}, and \ch{CH4}, were taken as test cases in
section~\ref{results-cc3-22}, using coordinates provided by \citeauthor{Kristiansen2022}~\cite{Kristiansen2022}
The coordinates of the \ch{H2} dimer for the $G'$
tensor calculation in section~\ref{results-cc3-23} can be found in the
dictionary of molecular structures of PyCC. All coordinates are also
available in the Supplementary Information (SI).

%% file: results.tex
\captionsetup[subfigure]{font={small}, skip=1pt, margin=-0.1cm, singlelinecheck=false}
\section{Results and Discussion} \label{results_cc3} 
\subsection{Computational Cost of the RT-CC3 Method} \label{results-cc3-1}
The CC3 method scales as ${\cal O}(N^{7})$, making it significantly more expensive than
the CCSD method. In the implementation, techniques including factorization,
reordering, and memory management need to be considered to improve efficiency,
while the scaling remains unchanged. 
This is standard practice in quantum chemistry and is expected when dealing with such steeply scaling methods, and has been extensively
discussed with respect to the CC3 method, in particular, in the literature.\cite{Koch1997,Christiansen1995CC3,Paul2020}
Taking the contribution from triples to
the $\hat{\Lambda}_{1}$ equation shown in Eq.~(\ref{eq:cc3-y1-exp}) as an
example, several adjustments can be made to accelerate the calculation. For
contractions involving three tensors, an intermediate consisting of two tensors
is calculated first to avoid a $N_{O}^{4}N_{V}^{4}$ contraction. The selection of the
two tensors in the initial step may also affect efficiency. For instance, we
can rewrite the third term in Eq.~(\ref{eq:cc3-y1-exp}) in two alternatives:
\begin{equation}
\sum\limits_{\substack{jkl\\bcd}}t_{jkl}^{bcd}L_{ij}^{ab}\lambda_{cd}^{kl} = \sum_{klcd}Z_{ikl}^{acd}\lambda_{cd}^{kl},
\end{equation}
where
\begin{equation}
Z_{ikl}^{acd}=\sum_{jb}t_{jkl}^{bcd}L_{ij}^{ab},
\end{equation}
or
\begin{equation}
\sum\limits_{\substack{jkl\\bcd}}t_{jkl}^{bcd}L_{ij}^{ab}\lambda_{cd}^{kl} = \sum_{jb}Z_{j}^{b}L_{ij}^{ab},
\end{equation}
where
\begin{equation}
Z_{j}^{b} = \sum_{klcd}t_{jkl}^{bcd}\lambda_{cd}^{kl}.
\end{equation}
The former approach results in a scaling of ${\cal O}(N^{6})$, whereas the intermediate
approach requires a contraction that scales at ${\cal O}(N^{8})$. The latter approach
results in a scaling of $N^{4}$ with an intermediate contraction of ${\cal O}(N^{6})$,
making it the favorable way to calculate this specific term. For the second
term in Eq.~(\ref{eq:cc3-y1-exp}), it can be rewritten as
\begin{equation}
\sum\limits_{\substack{jkl\\bcd}}t_{jkl}^{bcd}\bra{kl}ab \rangle \lambda_{cd}^{ij}=\sum_{jcd}Z_{j}^{acd}\lambda_{cd}^{ij},
\label{eq:cc3-result-l1-term211}
\end{equation}
where
\begin{equation}
Z_{j}^{acd} =\sum_{klb} t_{jkl}^{bcd}\bra{kl}ab \rangle,
\label{eq:cc3-result-l1-term212}
\end{equation}
or
\begin{equation}
\sum\limits_{\substack{jkl\\bcd}}t_{jkl}^{bcd}\bra{kl}ab \rangle \lambda_{cd}^{ij}=\sum_{klb}Z_{ikl}^{b}\bra{kl}ab \rangle,
\end{equation}
where
\begin{equation}
Z_{ikl}^{b} = \sum_{jcd}t_{jkl}^{bcd} \lambda_{cd}^{ij}.
\end{equation}
In this case, the two alternatives share the same scaling of ${\cal O}(N^{7})$ for the
contraction and ${\cal O}(N^{5})$ for the calculation of the intermediates. The former
factorization results in a scaling of $N_{O}^{3}N_{V}^{4}$, while the latter one
results in a scaling of $N_{O}^{4}N_{V}^{3}$. Following the same approach as was done
for the third term in the equation, the latter method should be preferable
since $N_{V}$ is usually larger than $N_{O}$ and grows faster when a larger basis set
is used. However, it is important to note that the intermediate $Z_{j}^{acd}$ in
Eqs.~(\ref{eq:cc3-result-l1-term211}) and~(\ref{eq:cc3-result-l1-term212}) does
not contain any $\hat{\Lambda}$ amplitudes, and thus it can be calculated before the
iterations and only needs to be computed once during the ground state
calculation. Similar considerations are taken into account for the other terms
in the CC3 equations as well.

Another computationally expensive step in the RT-CC3 method is the calculation
of the occupied-occupied block of the one-electron density, as shown in
Eq.~(\ref{eq:cc3-dij}). This calculation involves the contraction of the
$\hat{T}_{3}$ and $\hat{\Lambda}_{3}$ amplitudes, which only differ in one
index corresponding to the occupied orbital. Nested loops over virtual orbitals
are required for this calculation. For the density matrix elements, we choose
to implement a two-layer nested loop over virtual orbitals, considering it as a
four-index quantity for the triples with two fixed virtual orbitals. This is
preferred over a three-index quantity approach with three fixed virtual
orbitals. The contraction can be written as
\begin{equation}
\sum_{klc}\Omega_{ilk}^{c}\Omega_{c}^{jlk} \rightarrow D_{ij}
\end{equation}
for a certain pair of virtual orbitals $a,b$. Reducing the number of loops over
virtual orbitals accelerates the calculation of the one-electron density
substantially. For the ground state calculation, the density needs to be
calculated only once after the amplitudes converge from the iterations.
However, for RT simulations the acceleration of the density calculation is particularly important 
because it is carried out in every time step to obtain the corresponding time-dependent properties.

In addition to the above, the permutational symmetry of the amplitudes shown in
Eqs.~(\ref{eq:cc3-pijab}) and~(\ref{eq:cc3-pijkabc}), as well as the
permutational symmetry of the integrals, are facilitated in both derivation and
implementation. Identical terms that only differ in ordering need to be
identified to avoid repeated calculation with a polynomial scaling. Regarding
the triples, the amplitudes contracted with the same integral or other
amplitudes should be reordered first. For instance, in
Eq.~(\ref{eq:cc3-x1-exp}), $\hat{T}_{3}$ amplitudes contribute to $\hat{T}_{1}$
amplitudes by contracting with two-electron integrals. Two distinct triples are
required in the contraction. Instead of calculating two $\hat{T}_{3}$
amplitudes individually, the amplitudes are reordered so that they share the
same set of occupied orbitals. Given the known $t_{ijk}^{abc}$ with a fixed set
of $i,j,k$, $t_{ijk}^{cba}$ can be obtained simply by swapping the first and
third axis of the 3-index quantity. As noted by \citeauthor{Paul2020},~\cite{Paul2020}
the computational time for reordering can be
significant depending on the system size and the hardware used for the
calculation. Nevertheless, the calculation of an additional set of triple amplitudes is
still much more expensive and thereby dominant in the computational cost in our
implementation. Additionally, it is worth noting that the permutational
symmetry of the $T_{1}$-transformed integrals is reduced relative to the untransformed integrals: swapping both pairs of
indices in the bra and ket, \textit{viz.},
\begin{equation}
\langle pq\tilde{|} rs \rangle = \langle qp \tilde{|} sr \rangle.
\end{equation}

% Table1-CC3-timing
\begin{table} 
    \centering
	\caption{Performance comparison of RT-CC3/cc-pVDZ calculations for
	water clusters using different hardwares and precisions:
	double-precision on the CPU (CPU-dp), single-precision on the CPU
	(CPU-sp), double-precision on the GPU (GPU-dp), and single-precision on
	the GPU (GPU-sp). Timings (first four columns) are reported in seconds
	as per-step averages over five time steps. The final three columns
	indicate speed-ups, calculated as ratios of timings for each case.}
    \begin{tabular}{c|ccccccc}
       \textrm{Water Cluster} & $t_\textrm{CPU-dp}$ &  $t_\textrm{CPU-sp}$ & $t_\textrm{GPU-dp}$ &
$t_\textrm{GPU-sp}$ & $\frac{t_\textrm{CPU-dp}}{t_\textrm{GPU-dp}}$ & $\frac{t_\textrm{CPU-dp}}{t_\textrm{CPU-sp}}$ & 
$\frac{t_\textrm{GPU-dp}}{t_\textrm{GPU-sp}}$ \\ \hline
       \textrm{Monomer} & 16.105 & 11.192 & 18.511 & 18.661 & 0.86980 & 1.4390 & 0.99196 \\ 
       \textrm{Dimer} & 814.94 & 410.92 & 256.95 & 259.31 & 3.1716 & 1.9832 & 0.99090 \\
       \textrm{Trimer} & 10743 & 5364.1 & 806.52 & 768.49 & 13.320 & 2.0028 & 1.0495 \\
       \textrm{Tetramer} & & & 2455.7 & 1981.8 & & & 1.2391
    \end{tabular}
    \label{tab:cc3-gpu-cpu}
\end{table}

To assess the performance of our RT-CC3 implementation, we determined the
computational time for each time step, as shown in
Table~\ref{tab:cc3-gpu-cpu} for the water monomer, dimer, trimer, and tetramer.
Using the cc-pVDZ basis set, a single water molecule has 5 occupied orbitals
($N_{O}$) and 19 virtual orbitals ($N_{V}$). Each calculation was conducted exclusively
on a single node to ensure consistency in computational resources. All
contractions were done on either a CPU or a GPU.

\ZW{When transitioning from the monomer to the trimer, the system size increases by
a factor of three, theoretically causing the computational time to rise by a factor
of $3^7 \approx 2200$. As shown in the table, the CPU-dp calculation for the water trimer
takes approximately $3^{5.92}$ times longer than the monomer, while the running
time of the CPU-sp calculation increases by around $3^{5.62}$. For the GPU
calculations, the increase from the monomer to the trimer is approximately
$3^{3.44}$ for the double-precision calculation and $3^{3.38}$ for the
single-precision case. Furthermore, the GPU-dp calculation for the tetramer
takes about $4^{3.53}$ times longer than the monomer, while the
single-precision calculation takes approximately $4^{3.36}$ times longer. As
the system size continues to grow, the scaling will eventually reach ${\cal O}(N^{7})$ as
defined.}

It is evident that the application of single-precision does not achieve the ideal doubling
of the calculation speed, especially for GPU implementations. Nevertheless, the
speedup from CPU-sp becomes more noticeable as the system size increases, and a
discernible speedup emerges for GPU-sp calculations when the system size
reaches 96 molecular orbitals. Additionally, a considerable speedup was
attained from the GPU implementation overall. For the water trimer, the GPU-dp
calculation is 16 times faster than the CPU-dp calculation. We anticipate
further speedups from GPUs for even larger systems until a memory limitation is
encountered. It is also worth mentioning that our GPU implementation utilizes the
speedup of tensor contractions on GPUs, however, it is not fully optimized
in terms of memory allocation, parallelization, and other factors.

\subsection{Optical Properties}\label{results-cc3-2}
\subsubsection{Absorption Spectrum}\label{results-cc3-21}
%Fig1-RTCC3-AbsorptionSpectrum
\begin{figure}
    \centering
    \includegraphics[angle=0, scale=0.43]{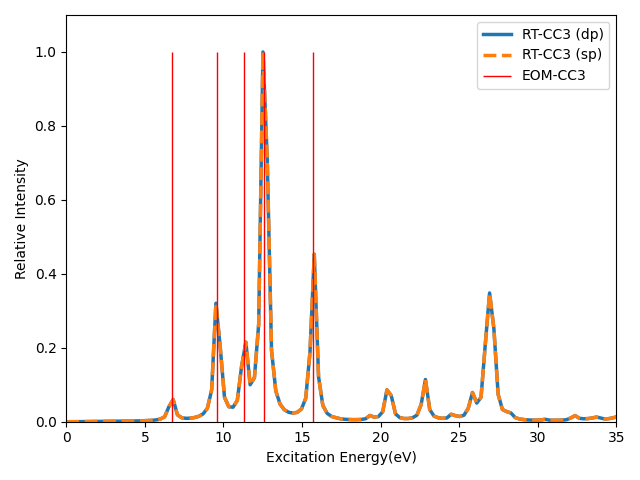}
    \caption{RT-CC3/cc-pVDZ linear absorption spectrum of \ch{H2O} with vertical
    lines indicating the corresponding EOM-CC3/cc-pVDZ excitation energies.}
    \label{fig:abs-water}
\end{figure}
To assess the stability and accuracy of the RT-CC3 implementation, we initially
calculated the linear absorption spectrum using the procedure outlined in
section~\ref{theory-cc3-21}. To generate a broadband spectrum, a thin Gaussian
pulse was applied. The absorption spectrum was computed for both single- and
double-precision arithmetics, as depicted in Fig.~\ref{fig:abs-water}. It has
been demonstrated that single-precision is sufficient for calculating the
absorption spectrum using RT-CCSD in our previous work.\cite{Wang2022}
Similarly, in this specific test case, no significant distinction between
single- and double-precision results is discernible in the RT-CC3 outcomes. For
the EOM-CC3/cc-pVDZ calculation, only excitation energies are attainable from
Psi4, while the corresponding oscillator strengths remain unavailable. For
illustrative purposes, the `height' of the stick spectra is chosen to be $1$
to enable convenient visualization of the position of each state.
However, this choice does not convey any information about the probability of the
corresponding transition. Through this comparison, we can ascertain that the
RT-CC3 method aligns well with the EOM-CC3 method.

\subsubsection{Dynamic Polarizabilities and First Hyperpolarizabilities}\label{results-cc3-22}
\begin{figure}
    \centering
    \includegraphics[angle=0, scale=0.43]{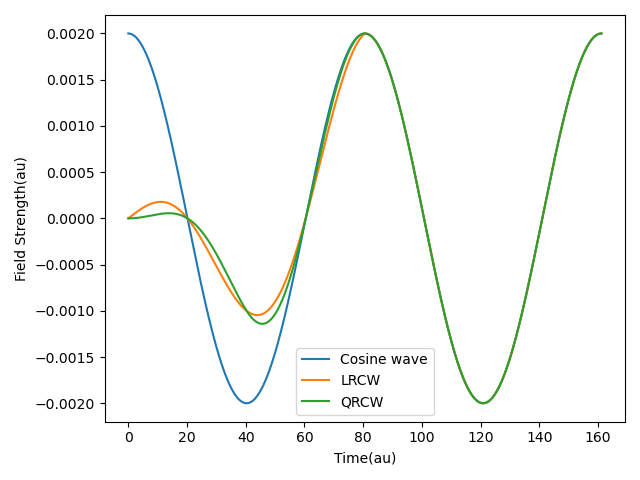}
    \caption{LRCW and QRCW of two optical cycles. Both of the RCWs have one
    ramped cycle following a cycle with a regular cosine wave. The frequency
    and the field strength are 0.078 au and 0.002 au, respectively.}
    \label{fig:rcw}
\end{figure}

As demonstrated in Ref.~\citenum{Ofstad2023}, the QRCW is favorable for
extracting optical properties as it has a smoother switch-on compared to the
LRCW or a simple oscillatory field without ramping. Fig.~\ref{fig:rcw}
illustrates that both LRCW and QRCW have significantly smaller amplitudes at
the initial stages of the simulation compared to the regular cosine wave. Compared
to LRCW, the
QRCW curve exhibits a more gradual increase during the first 20 au and more closely follows
the $\cos$ curve during the final 20 au of the ramping stage.
We apply both the LRCW and the QRCW to showcase the effect
of ramping.

%Fig2-RTCC3-RCW

\begin{figure}
     \centering
     \begin{subfigure}{0.47\textwidth}
         \centering
         \includegraphics[width=\textwidth]{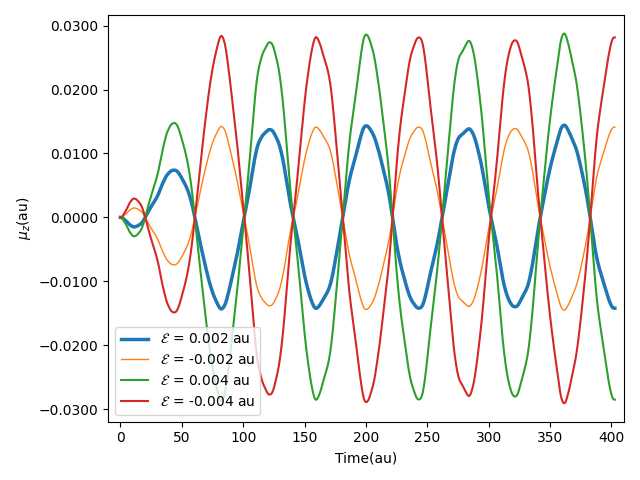}
     \end{subfigure}
     \hfill
     \begin{subfigure}{0.47\textwidth}
         \centering
         \includegraphics[width=\textwidth]{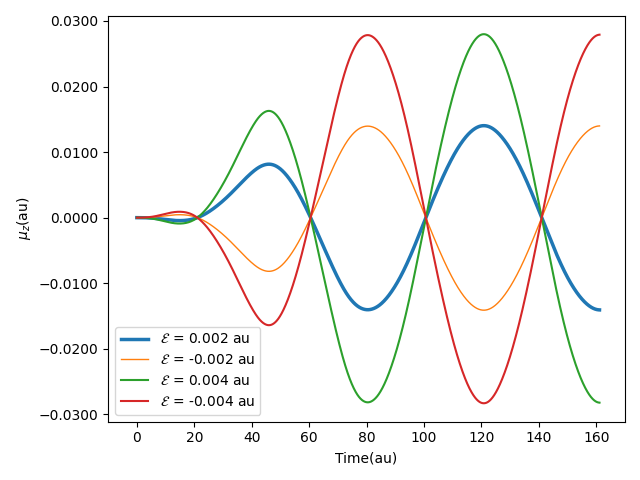}
     \end{subfigure}
     \vfill
     \begin{subfigure}{0.47\textwidth}
         \centering
         \includegraphics[width=\textwidth]{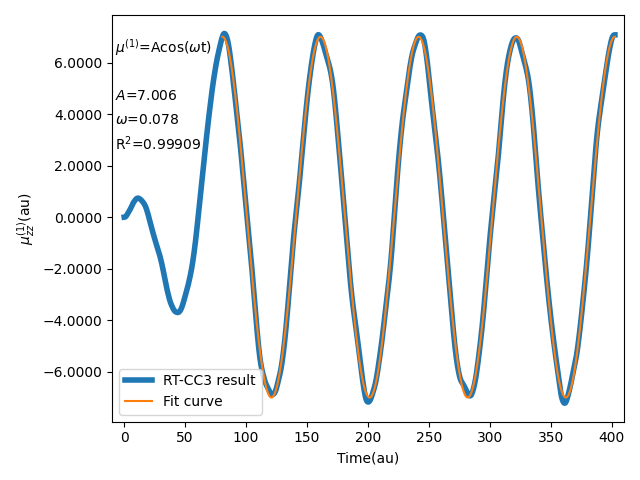}
     \end{subfigure}
     \hfill
     \begin{subfigure}{0.47\textwidth}
         \centering
         \includegraphics[width=\textwidth]{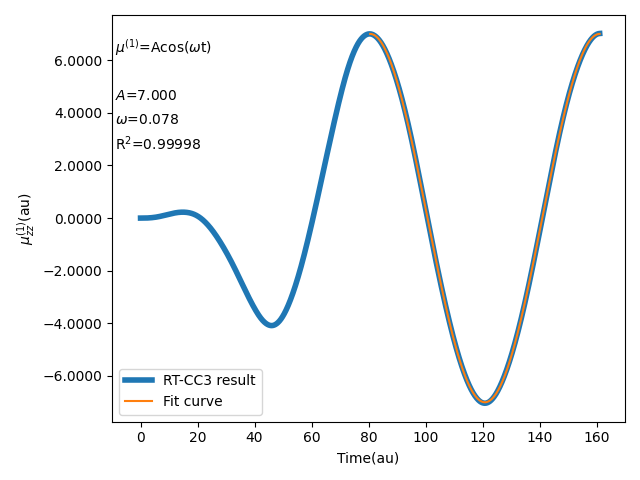}
     \end{subfigure}
     \vfill
     \begin{subfigure}{0.47\textwidth}
         \centering
         \includegraphics[width=\textwidth]{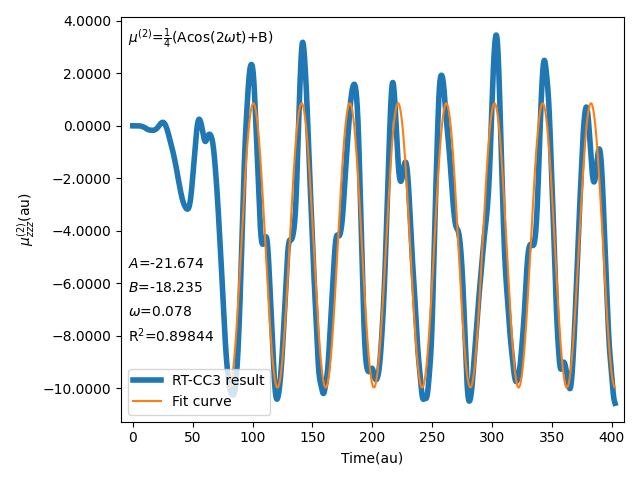}
     \end{subfigure}
     \hfill
     \begin{subfigure}{0.47\textwidth}
         \centering
         \includegraphics[width=\textwidth]{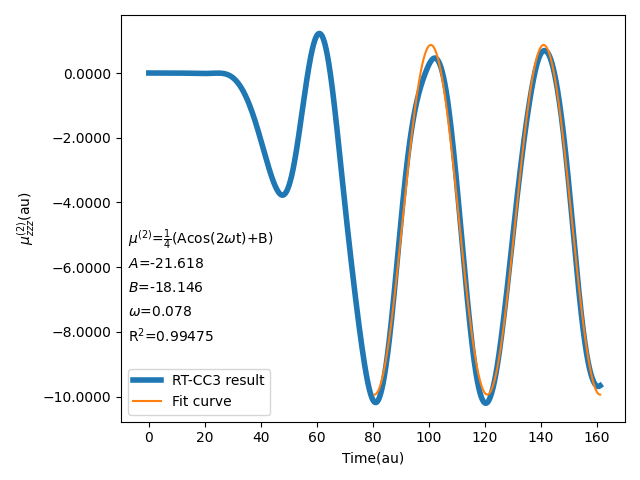}
     \end{subfigure}
     \caption{RT-CC3/cc-pVDZ (dp) results for the $z$-component of the induced dipole moment of \ch{H2O} from simulations
     with field strengths $\pm 0.002$ au and $\pm 0.004$ au.
     The left column displays the LRCW results with $n_r=1$ and $n_p=4$
     for the total dipole moment (top), the first- (middle) and second-order (bottom) dipole moments,
     including the curves obtained by fitting. The right column showcases the QRCW results with $n_r=n_p=1$.}
     \label{fig:pol-hyp-fit}
\end{figure}

Dynamic polarizabilities and first hyperpolarizabilities of \ch{H2O} at the
level of CCSD and CC3 are calculated using finite-difference methods. The
LRCW simulation spans five optical cycles, with the first cycle
reserved for linear ramping. In contrast, the QRCW simulation encompasses only two
optical cycles, again with ramping applied in the first cycle. The calculations are
conducted using both single-precision (sp) and double-precision (dp)
arithmetic. A representative result of RT-CC3/cc-pVDZ (dp) for \ch{H2O} is
depicted in Fig.~\ref{fig:pol-hyp-fit} to elucidate the procedure for obtaining
polarizabilities and first hyperpolarizabilities. 

From the fitted curve of the time trajectory of the first-order dipole moment,
the corresponding polarizability component can be calculated as the amplitude
of the curve. As depicted in Fig.~\ref{fig:pol-hyp-fit}, the values of
$\alpha_{zz}$ at $\omega=0.078$ au are \ZW{7.006 au and 7.000 au}, respectively, when
utilizing the LRCW and QRCW. Regarding the first hyperpolarizabilities, the
time trajectory of the second-order dipole moment is fitted into a cosine
curve, determining the amplitude $A$ and the phase $B$, which represent the
hyperpolarizabilities associated with SHG and OR, respectively. The quality of
the curve fitting is assessed using the R$^{2}$ value. As shown in
Fig.~\ref{fig:pol-hyp-fit}, a well-fitting curve is characterized by an R$^{2}$
value close to one, whereas a relatively inadequate fitting due to an
irregular-shaped second-order dipole trajectory is indicated by an R$^{2}$
value as low as \ZW{0.89839}. The summarized results are presented in
Tables~\ref{tab:polar},~\ref{tab:hyp-shg} and~\ref{tab:hyp-or}.
% Fig3: RT-CC3-pol-hyp-fit

To assess the performance of different simulations, three criteria are
evaluated: (1) accuracy compared to linear response (LR) CC, (2) R$^{2}$ value,
and (3) simulation length. A method capable of delivering accurate results from
a relatively short simulation, along with a curve fitting that yields a high
R$^{2}$ value, is the preferable choice. We employ the percentage error to
quantify accuracy, which is calculated using the following formula:
\begin{equation}
\textrm{Percentage Error} = \left| \frac{x - x_{0}}{x_{0}} \right| \times 100\%,
\end{equation}
where $x$ is the measured value and $x_{0}$ is the reference value. 
% Table2-polarizabilities
\begin{table}
  \centering
  \caption{RT-CCSD/cc-pVDZ and RT-CC3/cc-pVDZ polarizabilities (in atomic
  units) of \ch{H2O} at 582 nm from simulations with linear ramped
  continuous wave (LRCW) or quadratic ramped continuous wave (QRCW) fields.
  Reference values from LR-CCSD and LR-CC3 calculations using CFOUR are
  provided for comparison. The R$^{2}$ values, indicating the quality of curve
  fitting, are presented in the last three columns.}
  \begin{tabular}{c|c|ccc|ccc}
                                        & \textrm{Method} & $\alpha_{xx}$ & $\alpha_{yy}$ & $\alpha_{zz}$ 
                                          & R$^{2}_{\alpha_{xx}}$ & R$^{2}_{\alpha_{yy}}$ & R$^{2}_{\alpha_{zz}}$\\
                                          \hline
   & \textrm{LR-CCSD} & 3.182 & 10.549 & 7.017 & & &\\  
    \hline                                     
 \textrm{LRCW}  & \textrm{RT-CCSD\ (dp)} & 3.183 & 10.549 &  7.019 & 0.99980 & 0.99994 & 0.99909 \\
                           & \textrm{RT-CCSD\ (sp)} &  3.183 & 10.549 & 7.019 & 0.99980 & 0.99994 & 0.99909\\
    \cline{1-8}
  \textrm{QRCW} & \textrm{RT-CCSD\ (dp)} & 3.182 & 10.549 &  7.014 & 0.99999 & 0.99994 & 0.99998 \\
                           &\textrm{RT-CCSD\ (sp)} &  3.182 & 10.549 & 7.014 & 0.99999 & 0.99994 & 0.99998\\
    \hline\hline
   & \textrm{LR-CC3} & 3.164 & 10.581 & 7.007 & \\
    \hline
    \textrm{\ZW{LRCW}} &\textrm{RT-CC3\ (dp)} & 3.166 & 10.577 & 7.006 & 0.99981 & 0.99993 & 0.99909 \\
                             &\textrm{RT-CC3\ (sp)} & 3.166 & 10.577 & 7.006 & 0.99981 & 0.99993 & 0.99909 \\
    \cline{1-8}
    \textrm{\ZW{QRCW}} &\textrm{RT-CC3\ (dp)} & 3.164 & 10.576 & 7.001 & 0.99999 & 0.99994 & 0.99998 \\
                             &\textrm{RT-CC3\ (sp)} & 3.164 & 10.576 & 7.001 & 0.99999 & 0.99994 & 0.99998 \\
  \end{tabular}
  \label{tab:polar}
\end{table}

For polarizabilities, the single- and double-precision calculations yield
identical results up to three decimal places, with the same R$^{2}$ values
accurate for five decimal places. Minor discrepancy can be observed when
comparing to the LRCW  and the QRCW results. In RT-CCSD simulations, the LRCW
results exhibit a $0.03\%$ error in $\alpha_{xx}$ and a $0.03\%$ error in
$\alpha_{zz}$ , while the QRCW results show a $0.04\%$ error in $\alpha_{zz}$.
\ZW{In the case of RT-CC3 simulations, the LRCW results show a $0.06\%$ error in $\alpha_{xx}$, a $0.04\%$ error in
$\alpha_{yy}$, and a $0.01\%$ error in $\alpha_{zz}$ , while the QRCW results
indicate a $0.05\%$ error in $\alpha_{yy}$
and a $0.09\%$ error in $\alpha_{zz}$. Both of the two ramped continuous waves
yield errors well below $0.1\%$.} Importantly, QRCW requires less simulation
time compared to LRCW and offers a slightly better curve fitting. As a
result, the QRCW is the preferred choice, and this conclusion applies to both
RT-CCSD and RT-CC3 simulations.
% Table3-hyperpolarizabilities (SHG)
\begin{table}
  \centering
  \caption{RT-CCSD/cc-pVDZ and RT-CC3/cc-pVDZ first hyperpolarizabilities (in
  atomic units) associated with second harmonic generation (SHG) of \ch{H_{2}O}
  at 582 nm obtained from simulations with linear ramped continuous wave (LRCW)
  and quadratic ramped continuous wave (QRCW) fields. Reference values from
  LR-CCSD calculations using CFOUR are provided. The R$^{2}$ values, reflecting
  the quality of curve fitting, are displayed in the last three columns.}
  \begin{tabular}{c|c|ccc|ccc}
                                        &  \textrm{Method}  & $\beta_{zxx}$ & $\beta_{zyy}$ & $\beta_{zzz}$ 
                                          & R$^{2}_{\beta_{zxx}}$ & R$^{2}_{\beta_{zyy}}$ & R$^{2}_{\beta_{zzz}}$\\
                                          \hline
    & \textrm{LR-CCSD} & -4.091 & -35.441 & -22.423 & & &\\                 
    \hline                      
     \textrm{LRCW} & \textrm{RT-CCSD\ (dp)} & -4.311 & -35.694 &  -22.485 & 0.93021 & 0.54362 & 0.89911 \\
                              & \textrm{RT-CCSD\ (sp)} & -4.298 & -35.707 & -22.482 & 0.92954 & 0.54195 & 0.89921 \\
    \hline
     \textrm{QRCW} & \textrm{RT-CCSD\ (dp)} & -4.053 & -35.469 &  -22.435 & 0.99892 & 0.97345 & 0.99488 \\
                              & \textrm{RT-CCSD\ (sp)} & -3.987 & -35.520 & -22.447 & 0.99169 & 0.97266 & 0.99480 \\
    \hline\hline
    \textrm{\ZW{LRCW}} & \textrm{RT-CC3\ (dp)} & -4.081 & -35.675 & -21.674 & 0.92891 & 0.59673 & 0.89844 \\
                             & \textrm{RT-CC3\ (sp)} & -4.095 & -35.686 & -21.667 & 0.92874 & 0.59542 & 0.89839 \\
      \hline
     \textrm{\ZW{QRCW}} & \textrm{RT-CC3\ (dp)} & -3.828 & -35.414 & -21.618 & 0.99887 & 0.97110 & 0.99475 \\
                             & \textrm{RT-CC3\ (sp)} & -3.820 & -35.379 & -21.611 & 0.99273 & 0.97154 & 0.99474 \\
  \end{tabular}
  \label{tab:hyp-shg}
\end{table}
% Table4-hyperpolarizabilities (SHG)-error
\begin{table}
  \centering
    \caption{Percentage errors of RT-CCSD/cc-pVDZ first hyperpolarizabilities
    (au) associated with SHG of \ch{H2O} at 582 nm from calculations using
    LRCW and QRCW.}
  \begin{tabular}{c|c|ccc}
                                        &  \textrm{Method}  & $\beta_{zxx}$ & $\beta_{zyy}$ & $\beta_{zzz}$ \\
                                          \hline                   
     \textrm{LRCW} & \textrm{RT-CCSD\ (dp)} & 5.38\% & 0.71\% &  0.28\%  \\
                              & \textrm{RT-CCSD\ (sp)} & 5.06\% & 0.75\% & 0.26\%  \\
    \hline
     \textrm{QRCW} & \textrm{RT-CCSD\ (dp)} & 0.93\% & 0.08\% &  0.05\%  \\
                              & \textrm{RT-CCSD\ (sp)} & 2.59\% & 0.22\% & 0.11\%  \\
     \end{tabular}
      \label{tab:hyp-shg-error}
\end{table}
% Table5-hyperpolarizabilities (OR)
\begin{table}
  \centering
    \caption{RT-CCSD/cc-pVDZ and RT-CC3/cc-pVDZ first hyperpolarizabilities (in
    atomic units) associated with optical rectification (OR) of \ch{H2O} at
    582 nm obtained from simulations with linear ramped continuous wave (LRCW)
    and quadratic ramped continuous wave (QRCW) fields. Reference values from
    LR-CCSD calculations using CFOUR are provided. The R$^{2}$ values,
    indicating the quality of curve fitting, are displayed in the last three
    columns.}
  \begin{tabular}{c|c|ccc|ccc}
                                        &  \textrm{Method}  & $\beta_{zxx}$ & $\beta_{zyy}$ & $\beta_{zzz}$ 
                                          & R$^{2}_{\beta_{zxx}}$ & R$^{2}_{\beta_{zyy}}$ & R$^{2}_{\beta_{zzz}}$\\
                                          \hline
   & \textrm{LR-CCSD} & -4.488 & -30.485 & -18.830 & & &\\  
    \hline                                     
    \textrm{LRCW} & \textrm{RT-CCSD\ (dp)} & -4.532 & -30.568 &  -18.927 & 0.93021 & 0.54362 & 0.89911 \\
                              &   \textrm{RT-CCSD\ (sp)} &  -4.579 & -30.624 & -18.977 & 0.92954 & 0.54195 & 0.89921 \\
    \hline
    \textrm{QRCW} & \textrm{RT-CCSD\ (dp)} & -4.481 & -30.513 &  -18.848 & 0.99892 & 0.97345 & 0.99488 \\
                              &   \textrm{RT-CCSD\ (sp)} &  -4.445 & -30.733 & -18.918 & 0.99169 & 0.97266 & 0.99480 \\
    \hline\hline
     \textrm{\ZW{LRCW}} & \textrm{RT-CC3\ (dp)} & -4.292 & -30.516 & -18.235 & 0.92891 & 0.59673 & 0.89844 \\
                               &   \textrm{RT-CC3\ (sp)} & -4.289 & -30.529 & -18.210 & 0.92874 & 0.59542 & 0.89839 \\
      \hline
      \textrm{\ZW{QRCW}} & \textrm{RT-CC3\ (dp)} & -4.240 & -30.415 & -18.146 & 0.99887 & 0.97110 & 0.99475 \\
                               &   \textrm{RT-CC3\ (sp)} & -4.312 & -30.433 & -18.128 & 0.99273 & 0.97154 & 0.99474 \\
  \end{tabular}
  \label{tab:hyp-or}
\end{table}
% Table6-hyperpolarizabilities (OR)-error
\begin{table}
  \centering
    \caption{Percentage errors of RT-CCSD/cc-pVDZ first hyperpolarizabilities
    (au) associated with OR of \ch{H2O} at 582 nm from calculations using
    LRCW and QRCW.}
  \begin{tabular}{c|c|ccc}
                                        &  \textrm{Method}  & $\beta_{zxx}$ & $\beta_{zyy}$ & $\beta_{zzz}$ \\
                                          \hline                   
     \textrm{LRCW} & \textrm{RT-CCSD\ (dp)} & 0.98\% & 0.27\% &  0.51\%  \\
                              & \textrm{RT-CCSD\ (sp)} & 2.03\% & 0.46\% & 0.78\%  \\
    \hline
     \textrm{QRCW} & \textrm{RT-CCSD\ (dp)} & 0.16\% & 0.09\% &  0.10\%  \\
                              & \textrm{RT-CCSD\ (sp)} & 0.96\% & 0.81\% & 0.47\%  \\
     \end{tabular}
  \label{tab:hyp-or-error}
\end{table}

For the first hyperpolarizabilities, we observe larger errors in RT-CCSD
results compared to LR-CCSD, as well as differences between single- and
double-precision results. This outcome is reasonable, considering that we are
calculating higher-order induced dipole moments. It is
important to note that the R$^{2}$ values in Tables~\ref{tab:hyp-shg}
and~\ref{tab:hyp-or} are identical. This is because the hyperpolarizabilities
associated with second harmonic generation (SHG) and optical rectification (OR)
are obtained using the same curve fitting process, with the field applied in a
specific direction.

Tables~\ref{tab:hyp-shg-error} and~\ref{tab:hyp-or-error} summarize the
percentage errors of hyperpolarizability elements obtained from RT-CCSD
calculations, compared to LR-CCSD. In Table~\ref{tab:hyp-shg-error}, the
largest error of $5.38\%$ occurs in $\beta_{zxx}$ from the RT-CCSD (dp)
calculation using LRCW. By switching to the QRCW, the error in $\beta_{zxx}$ is
reduced by $82.71\%$, to $0.93\%$. The percentage errors for other
elements are also substantially reduced by at least $48.81\%$. Moreover, the
R$^{2}$ values improve when using the QRCW, as seen in Table~\ref{tab:hyp-shg}.
Notably, for $\beta_{zyy}$ where the applied field is perpendicular to the
molecular plane, a less smooth trajectory of the second-order dipole moments
leads to lower R$^{2}$ values for the LRCW case. Using the QRCW recovers the
quality of curve fitting, with R$^{2}$ values exceeding $0.97$.

Regarding precision arithmetic, the double-precision calculation with the LRCW
outperforms the single-precision case for $\beta_{zyy}$, while showing slightly
worse results for $\beta_{zxx}$ and $\beta_{zzz}$. However, for calculations
using the QRCW, the single-precision arithmetic leads to larger errors for all
elements. Generally, a double-precision calculation should yield more accurate
and robust results because double-precision floating-point numbers are accurate
up to 15 digits, whereas single-precision numbers are accurate only up
to around seven digits. In our test case, when the LRCW is used, the major
error arises from the choice of ramping. This can be observed from the
relatively large overall error and the poor R$^{2}$ values. The difference
caused by the two different precision arithmetics is not as pronounced. Neither
of them produces sufficiently accurate results. However, when the QRCW is
employed, the percentage error is significantly reduced due to the more gradual
and smooth switch-on of the field, regardless of the chosen precision arithmetics.
Consequently, the lower precision arithmetic becomes the primary factor
contributing to the resulting error. This is evident in the last two rows of
Table~\ref{tab:hyp-shg-error}, where errors in single-precision calculations
are more than twice those in the double-precision calculations.

A similar analysis applies to the results of hyperpolarizabilities associated
with OR, as shown in Tables~\ref{tab:hyp-or} and~\ref{tab:hyp-or-error}. In
this case, the percentage errors originating from the LRCW are not as
significant as those observed in the case of hyperpolarizabilities associated
with SHG. Results from the double-precision calculations are consistently more
accurate than the single-precision results. Furthermore, the QRCW continues to
significantly enhance accuracy for each element, consistent with the trends
observed for hyperpolarizabilities associated with SHG.

In the context of RT-CC3 calculations, even though reference values are
unavailable for direct comparison, the impact of replacing the LRCW with the
QRCW is evident from the substantial increase in R$^{2}$ values. The excellent
R$^{2}$ values observed in the QRCW RT-CC3 calculations further
reinforce the notion that our implementation serves as a viable tool for
calculating dynamic polarizability and first hyperpolarizabilities at the CC3
level, given the limitations of available alternatives.

The results presented above demonstrate the capability of the RT-CC3 method for
calculating polarizabilities and first hyperpolarizabilities. Given the
approximated orbital relaxation with singles in CC3, it is worthwhile to
explore a comparison to orbital-optimized coupled cluster (OCC) methods where
the singles cluster operators are replaced by orbital rotations. As
an example, \citeauthor{Kristiansen2022}\cite{Kristiansen2022}
implemented real-time (RT) time-dependent
orbital-optimized second-order M{\o}ller-Plesset (TDOMP2) theory,\cite{Pathak2020}
which serves as a
second-order approximation to the time-dependent orbital-optimized coupled
cluster doubles (TDOCCD) method.\cite{Pedersen1999,Sato2018} TDOMP2 is further
compared to RT-CC2, which is a second-order approximation to RT-CCSD.
\citeauthor{Kristiansen2022} showed that while orbital optimization does not
significantly affect linear absorption spectra, it leads to a significant
improvement relative to RT-CC2 theory for polarizabilities and hyperpolarizabilities.
This observation also holds for complex-valued polarizabilities obtained in the
presence of a static uniform magnetic field.\cite{Ofstad2023_magnetic}

In addition to the TDOMP2 method, \citeauthor{Kristiansen2022} also developed the
time-dependent nonorthogonal OCCD (TDNOCCD) method,\cite{Pedersen2001,Kvaal2012} where the orbital rotation
is non-unitary, which is crucial for convergence to the FCI limit.\cite{Kohn2005,Myhre2018}
To assess the performance of RT-CC3 and TDNOCCD for
polarizabilities, several ten-electron systems are investigated using
double-precision calculations to mitigate errors stemming from low-precision
arithmetic. Table~\ref{tab:noccd} presents the TDNOCCD results computed using
the Hylleraas Quantum Dynamics (HyQD) software library\cite{HyQD}
and compares them with our RT-CC3 results. Reference values include
FCI values and LR results, with RT-CCSD results included for comparison. FCI
and LR-CC3 values for \ch{Ne} and \ch{HF} are obtained from
Ref.~\citenum{Larsen1999}. LR-CC3 values for other molecules are computed using
CFOUR. All RT simulations employ the QRCW as the applied field with $n_r=n_p=1$.

\input{table_rtcc3_vs_tdnoccd}

For \ch{Ne}, RT-CC3 exhibits good agreement with LR-CC3 and FCI, with errors of
at most $0.67\%$ for frequencies ranging from 0.1 au to 0.3 au. However, a
notable deviation from LR-CC3 and FCI results becomes apparent at a frequency of
0.4 au, which is closer to the resonance at 0.613 au. The accuracy of the
result at $\omega=0.5$ au is expected to be even lower, as indicated by the
comparison between LR-CCSD and RT-CCSD. In fact, the RT-CC3 result at
$\omega=0.5$ au is closer to the reference value, which is likely coincidental.
To assess the quality of
curve fitting, R$^{2}$ values are compared for different frequencies. \ZW{The
R$^{2}$ values for $\omega=0.3$ au, $\omega=0.4$ au, and $\omega=0.5$ au are
0.99651, 0.99318, and 0.98151, respectively.} These values decrease with higher
frequencies. While the result at $\omega=0.5$ au is ``accurate,'' it is somewhat
less reliable than the results for lower frequencies.

A similar trend is observed for \ch{HF}. RT-CC3 regains accuracy compared
to LR-CC3 and FCI at frequencies of 0.1 au and 0.2 au. At a frequency of 0.3
au, which is quite close to the resonance at 0.383 au, the accuracy decreases,
resulting in percentage errors of $8.13\%$ and \ZW{$2.09\%$} for $\alpha_{yy}$ and
$\alpha_{zz}$, respectively. \ZW{Corresponding $R^{2}$ values are 0.97622 and
0.97835, respectively.} The slightly larger error of $\alpha_{yy}$ is related to
the symmetry of \ch{HF}. The first excitation involves one of the lone pair
electrons of Fluoride and the $\sigma^{*}$ orbital. Since the lone pair
electrons align with the y-axis, the $\alpha_{yy}$ component is more relevant
to the transient and therefore exhibits a larger percentage error. The observed 
pattern in CC3 results aligns with that in the CCSD results.

For \ch{H2O}, selected frequencies are all well below the resonance at 0.277 au.
RT-CC3 values consistently align with LR-CC3, with only a $0.11\%$ error
observed in $\alpha_{zz}$ at $\omega=0.1\ au$. In the case of \ch{NH3},
RT-CC3 results match LR-CC3 values for all frequencies chosen, which are all
below the resonance at 0.236 au. The exception is $\alpha_{zz}$ at $\omega=0.1$ au
where a $0.93\%$ error occurs. This small discrepancy contrasts with the
agreement between LR-CCSD and RT-CCSD at the same frequency. Notably, 
the lone pair electrons of nitrogen that are significant to the lowest excitation 
level are aligned with the z-axis.

A discrepancy is also seen in the results for \ch{CH4} at the frequency of
0.2 au, whereas the resonance occurs at 0.38 au. RT-CC3 results show a $1.07\%$
deviation from LR-CC3, compared to a mere $0.15\%$ discrepancy in the CCSD
case. The corresponding R$^{2}$ values for these two less accurate results,
$\alpha_{zz}$ of \ch{NH3} and the polarizability of \ch{CH4}, are 0.99823
and 0.99574, respectively. These values are smaller than those of the other
polarizability values, which are all above 0.9999.

To further explore the differences between RT-CCSD and RT-CC3 in these cases,
we calculated two additional LR-CCSD/LR-CC3 polarizabilities at different
frequencies and performed polynomial regression with five data points. This
analysis reveals the relationship between increasing polarizabilities and
frequency. In addition to the values listed in Table~\ref{tab:noccd},
$\alpha_{zz}$ of \ch{NH3} is calculated at a frequency of 0.025 au, yielding
LR-CC3 and LR-CCSD results of 14.88 au and 14.90 au, respectively. At a
frequency of 0.085 au, $\alpha_{zz}$ values are 15.72 au and 15.74 au for
LR-CC3 and LR-CCSD, respectively. Two more frequencies, 0.0428 au and 0.15 au,
are selected for \ch{CH4}. RT-CC3 and RT-CCSD results at $\omega=0.0428\ au$
are 16.89 au and 16.91 au, respectively. At $\omega=0.15$ au, the
corresponding results are 18.20 au and 18.21 au, respectively.

The polarizability can be written as a Taylor expansion containing only even orders of
the frequency $\omega$,~\cite{Hirschfelder1964}
\begin{align}
    \alpha_{\beta\beta}(\omega) = \sum_{i=0}^\infty S_{\beta\beta}(-2i - 2) \omega^{2i},
  \label{eq:cauchy-moments}
\end{align}
which converges for frequencies below the first excitation energy.
The coefficients $S_{\beta\beta}(i)$ are oscillator-strength sum rules, also known as
Cauchy moments,~\cite{Hattig1997}
and contain a wealth of information about molecular properties.~\cite{Hirschfelder1964}
\citeauthor{Hattig1997} have studied
the Cauchy moments using LR-CCS, LR-CC2, and LR-CCSD theory.~\cite{Hattig1997}
Here, we were able to obtain the Cauchy moments using theories at the level of
CCSD and CC3, both LR theory and RT methods for comparison. The Taylor expansion was 
truncated after the third term ($\omega^4$) for both LR and RT results.

\begin{figure}
     \centering
     \begin{subfigure}{0.495\textwidth}
         \centering
         \includegraphics[width=\textwidth]{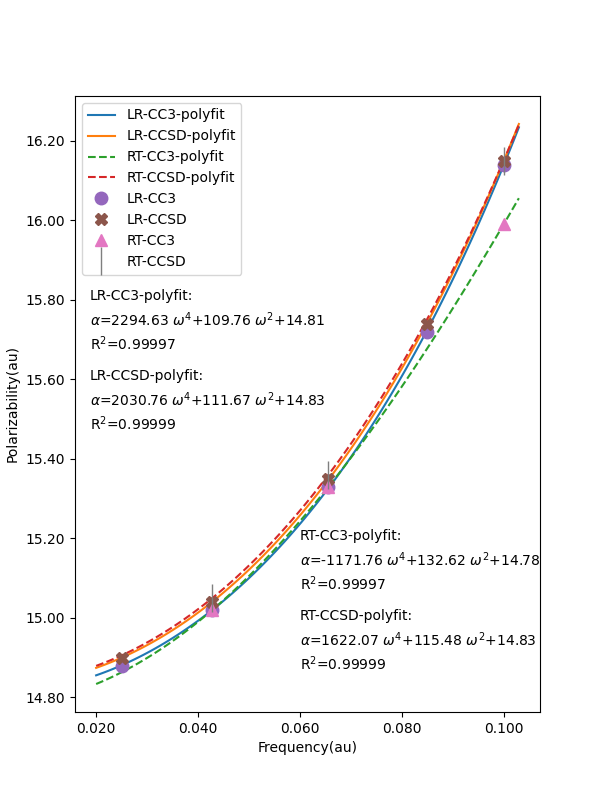}
     \end{subfigure}
     \hfill
     \begin{subfigure}{0.495\textwidth}
         \centering
         \includegraphics[width=\textwidth]{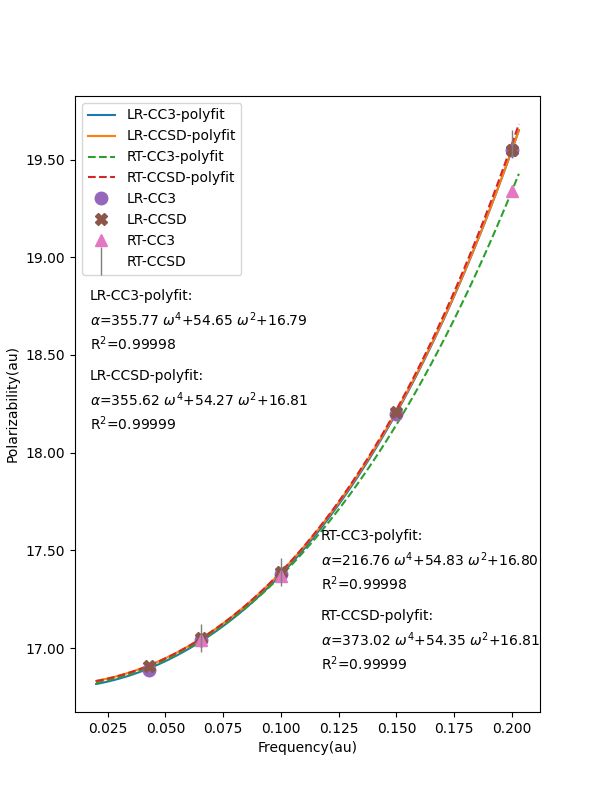}
     \end{subfigure}
     \caption{Dispersion of $\alpha_{zz}$ for \ch{NH3} (left) and
     polarizabilities for \ch{CH4} (right),
     calculated using CC3 and CCSD methods. Polynomial regression curves are
     depicted, along with the resulting functions and R$^{2}$ values as
     annotations on the figures.}
     \label{fig:polar-polyfit}
\end{figure}

As shown in Fig.~\ref{fig:polar-polyfit}, polynomial regression for LR-CCSD and
LR-CC3 results closely aligns with data points for all four data sets, with
R$^{2}$ values exceeding 0.9999. Moreover, the coefficients of $\omega^{4}$
from LR-CC3 data are larger than those from LR-CCSD data for both \ch{NH3} and
\ch{CH4}. The polynomial regression results suggest that CC3 polarizabilities
increase slightly faster with frequency compared to CCSD polarizabilities. The
impact of frequency moving towards the resonance on polarizability results
becomes evident earlier in the frequency range for CC3 compared to CCSD in
these test cases, potentially explaining the disagreement observed for
polarizability values of \ch{NH3} and \ch{CH4}. For the RT-CCSD and RT-CC3
results, the coefficients of $\omega^{4}$ from RT-CC3 data are smaller than
those from RT-CCSD data. The $R^{2}$ values are still close to $1.0$, however,
the regression is largely affected by the polarizabilities near resonance,
especially involving only three data points. The coefficients of $\omega^{4}$
deviates from LR results, while the ones of $\omega^{2}$ and $\omega^{0}$ does
not deviate as much. More data points in the frequency range that are away from
the resonance should resolve the deviation as the polarizabilites align well
with the ones from LR calculations. In the meantime, $S_{\beta\beta}(-2i-2)$
are obtained from the regression according to Eq.~(\ref{eq:cauchy-moments}).
To confirm the accuracy at low frequencies, the $S_{\beta\beta}(-2)$ coefficient 
is compared to the static polarizability. For \ch{NH3}, the static 
polarizabilities obtained from LR-CCSD and LR-CC3 are $14.83$ and $14.81$ a.u.\ 
respectively. The RT-CC3 result has a $0.2\%$ relative difference 
compared to its reference value, while the RT-CCSD error is effectively zero. For \ch{CH4}, the static polarizabilities 
obtained from LR-CCSD and LR-CC3 are $16.80$ and $16.79$ respectively. All the 
$S_{\beta\beta}(-2)$ results align well with the reference value with the 
relative difference smaller than $0.1\%$. We have shown that the method is 
capable of calculating Cauchy moments conveniently using theories at the level 
of CCSD and CC3, although more data points will be needed to obtain accurate 
values of Cauchy moments with larger $i$. 

% Fig4: polyfit-polar

The impact of orbital optimization is explored by comparing TDNOCCD with
RT-CCSD and higher levels of theory. As previously mentioned, TDNOCCD differs
from RT-CCSD by substituting singles with a non-unitary orbital rotation, where
the rotation parameters are time-dependent. Orbital-optimized CC methods have
advantages in multi-electron ionization dynamics, chemical bond breaking, response theory,
and more.\cite{Pedersen2001, Kvaal2012, Kohn2005, Myhre2018, Sato2018,
Pathak2020, Pathak2022, Ofstad2023} Explicit orbital optimization also enhances
the stability of real-time simulations when systems are subjected to strong
external fields, and the ground state no longer dominates because the
time-dependent reference determinant tends to be close or identical to the
Brueckner determinant.\cite{Kristiansen2020}

In our test cases, TDNOCCD is initially compared to RT-CCSD by assessing its
differences from LR-CCSD. The data presented in Table~\ref{tab:noccd} indicate
that TDNOCCD results exhibit relative differences ranging from $0.89\%$ to
$7.09\%$, with an average difference of $2.98\%$ across all frequencies and
molecules, compared to LR-CCSD. The most significant difference arises from the
polarizability of \ch{Ne} at $\omega=0.5\ au$. However, this TDNOCCD result is
actually closer to LR-CCSD than RT-CCSD for this specific value. Unlike the RT
methods, the frequency dependence of relative differences in TDNOCCD is not as
pronounced. It is evident that substantial differences are present not only in
high-frequency results but also in low-frequency outcomes, which are distant
from resonances. The primary factor contributing to this divergence between
TDNOCCD and RT-CCSD, in comparison with LR-CCSD, is the orbital optimization.

Next, TDNOCCD results and RT-CC3 are compared to LR-CC3. As documented in
Ref.~\citenum{Larsen1999}, LR-CC3 can be taken as a reference value considering
its high accuracy compared to FCI, although this choice gives a slight bias
towards methods without orbital optimization.
Theoretically, RT-CC3 should exactly
reproduce LR-CC3 (up to numerical differentiation, accuracy of the integrator,
etc.) by adding more ramping cycles. 
Except for a few cases (\ch{Ne} at $\omega=0.4$ au and $0.5$ au, \ch{HF} 
at $\omega=0.3$ au, \ch{NH3} at $\omega=0.1$ au, and \ch{CH4} at
$\omega=0.2$ au), RT-CC3 results match LR-CC3. TDNOCCD, however, deviates from
LR-CC3/RT-CC3 by at least $1.03\%$ and up to $3.41\%$, with an average
deviation of $2.14\%$. As these polarizability results are unaffected by
proximity to resonances, the deviation stems from the distinct treatments
of orbital optimization and the inclusion or exclusion of triples.

In cases where RT-CC3 exhibits significant percentage errors compared to
LR-CC3, TDNOCCD may be closer or farther from LR-CC3.
When
RT-CC3 overestimates polarizability of \ch{Ne} ($\omega=0.4$ au) and
$\alpha_{yy}$ of \ch{HF} ($\omega=0.3$ au) by $10.54\%$ and $8.13\%$,
respectively, TDNOCCD yields smaller values closer to LR-CC3 due to orbital optimization,
underestimating these polarizability values.
When RT-CC3
underestimates polarizabilities, the even smaller values from TDNOCCD result in
a larger difference from LR-CC3.

% G' tensor 
\subsubsection{$G'$ tensor and quadratic response function}\label{results-cc3-23}
% Fig5: RT-CC3-G' tensor
\begin{figure}
     \centering
     \begin{subfigure}{0.47\textwidth}
         \centering
         \includegraphics[width=\textwidth]{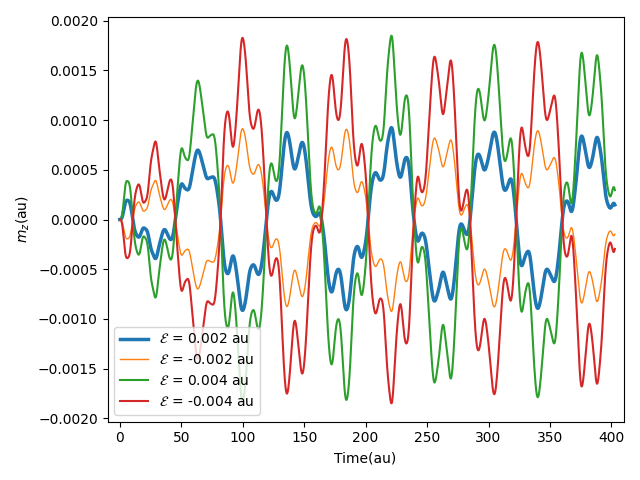}
     \end{subfigure}
     \hfill
     \begin{subfigure}{0.47\textwidth}
         \centering
         \includegraphics[width=\textwidth]{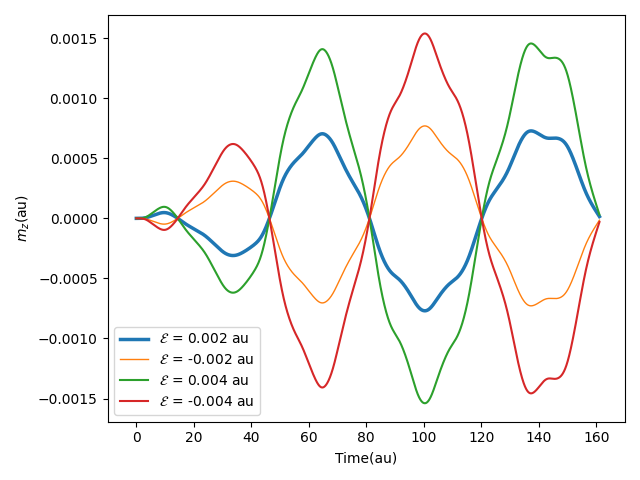}
     \end{subfigure}
     \vfill
     \begin{subfigure}{0.47\textwidth}
         \centering
         \includegraphics[width=\textwidth]{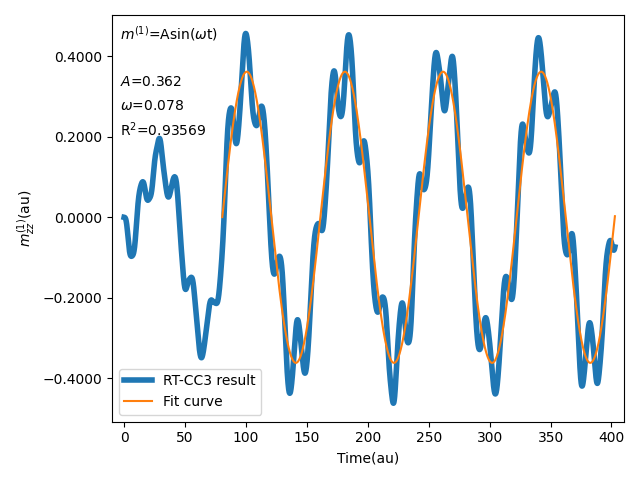}
     \end{subfigure}
     \hfill
     \begin{subfigure}{0.47\textwidth}
         \centering
         \includegraphics[width=\textwidth]{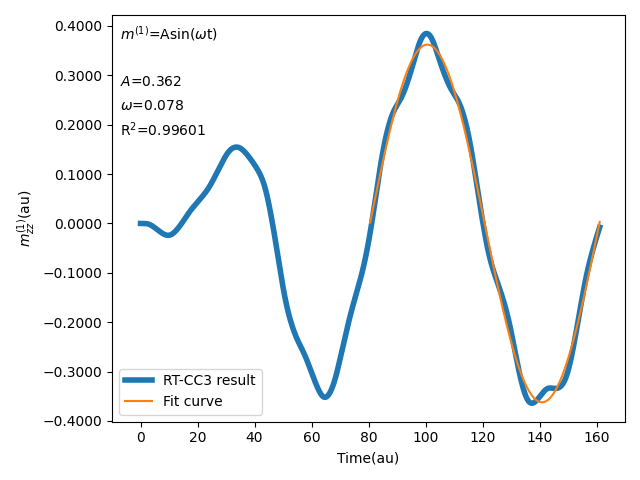}
     \end{subfigure}
     \vfill
     \begin{subfigure}{0.47\textwidth}
         \centering
         \includegraphics[width=\textwidth]{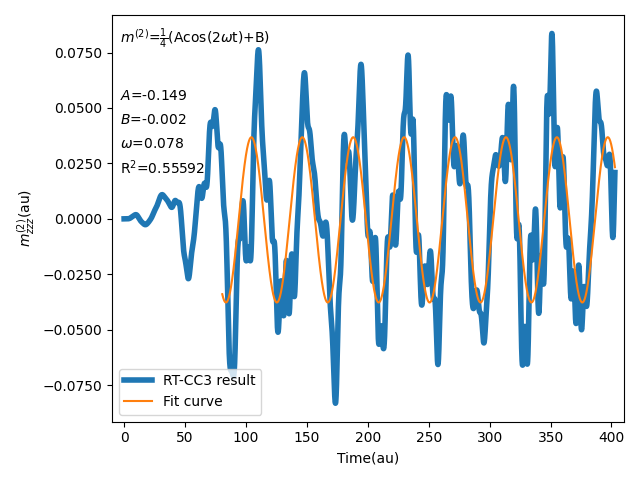}
     \end{subfigure}
     \hfill
     \begin{subfigure}{0.47\textwidth}
         \centering
         \includegraphics[width=\textwidth]{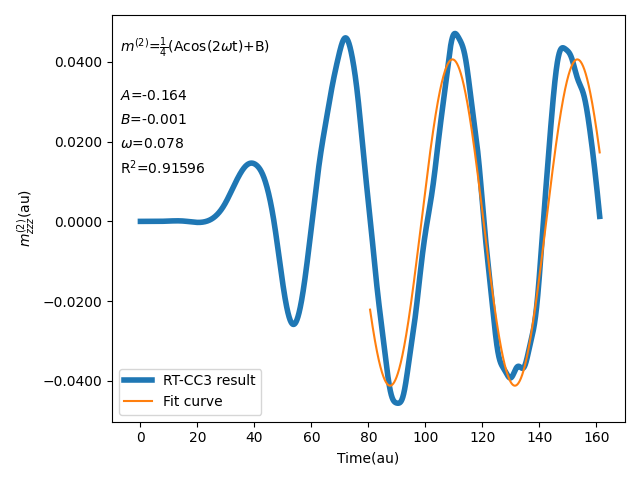}
     \end{subfigure}
     \caption{RT-CC3/cc-pVDZ (dp) results of \ch{H_{2}} dimer obtained from
     four simulations with field strengths of 0.002 au, -0.002 au, 0.004 au,
     and -0.004 au. The left column displays the LRCW results, including the
     $z$ component of the induced magnetic dipole moment and the corresponding
     first-order and second-order dipole moments with their fitted curve. The right column
     presents the QRCW results.}
     \label{fig:g-fit}
\end{figure}	
With our RT-CC implementation, we calculate the $G'$ tensor of the \ch{H_{2}}
dimer using RT-CCSD and RT-CC3, both in single- and double-precision
arithmetics. We employ both the LRCW and the QRCW for comparison purposes. As
illustrated in Fig.~\ref{fig:g-fit}, we utilize the induced magnetic dipole
moments from applied fields of varying strengths to compute the first-order
magnetic dipole moments through the finite difference method, similar to the
procedure for calculating polarizabilities. In this case, $G'_{zz}$ can be
obtained from the magnetic dipole moment induced by the field applied in the
$z$ direction, with its value represented by the amplitude of the fitted curve.

As per Table~\ref{tab:g-ccsd}, no distinction is observed between single- and
double-precision results in the RT-CCSD calculations. The $G'$ tensor elements
exhibit identical values for both the LRCW and the QRCW cases, with the
distinction lying solely in the R$^{2}$ values. Notably, the QRCW significantly
enhances curve fitting quality, aligning with the conclusion drawn in
section~\ref{results-cc3-22}. Upon examining the example results in
Fig.~\ref{fig:g-fit}, it is evident that the induced magnetic dipole moment
curves are less smooth compared to the induced electric dipole moment curves
discussed in the previous section. Particularly in the LRCW instance, a
curvilinear trajectory of the dipole is observed. Although this irregular shape
remains approximately periodic, it adversely affects curve fitting. A minor
distortion is observed in the QRCW example, which has a lesser impact on curve
fitting. Table~\ref{tab:g-cc3} presents the RT-CC3 results. Analogous to
RT-CCSD, the disparities between single- and double-precision results are
inconsequential. The QRCW enhances overall R$^{2}$ values and provides more
reliable outcomes. Hence, the RT-CC3 method combined with QRCW is a viable approach
for computing the $G'$ tensor and subsequently optical rotation.
% Table8-G' tensor
\begin{table}
  \centering
    \caption{RT-CCSD/cc-pVDZ $G'$ tensor elements (in atomic units) of H$_{2}$
    dimer at 582 nm obtained using single- and double-precision computations,
    with linear ramped continuous wave (LRCW) and quadratic ramped continuous
    wave (QRCW) fields. The R$^{2}$ values, indicating the quality of curve
    fitting, are displayed in the last three columns.}
  \begin{tabular}{c|c|c|ccc}
                                          &  $G'$ & \textrm{RT-CCSD(dp)} & R$^{2}_{x}$(dp) & R$^{2}_{y}$(dp)& R$^{2}_{z}$(dp)\\
                                          \hline
                \textrm{LRCW} & ($G'_{xx}$, $G'_{yx}$, $G'_{zx}$) & (-0.387, -0.097, 0.000) &   0.93726 & 0.93911 & 0 \\
                                          & ($G'_{xy}$, $G'_{yy}$, $G'_{zy}$) & ( 0.058,  0.013, 0.000) &   0.88177 & 0.86203 & 0 \\
                                          & ($G'_{xz}$, $G'_{yz}$, $G'_{zz}$) & ( 0.000,  0.000, 0.362) &   0 & 0 & 0.93669 \\   
                \hline
                                           &  $G'$ & \textrm{RT-CCSD(sp)} & R$^{2}_{x}$(sp) & R$^{2}_{y}$(sp)& R$^{2}_{z}$(sp)\\
                                          \hline
                \textrm{LRCW} & ($G'_{xx}$, $G'_{yx}$, $G'_{zx}$) & (-0.387, -0.097, 0.000) &   0.93727 & 0.93911 & 0 \\
                                          & ($G'_{xy}$, $G'_{yy}$, $G'_{zy}$) & ( 0.058,  0.013, 0.000) &   0.88179 & 0.86205 & 0 \\
                                          & ($G'_{xz}$, $G'_{yz}$, $G'_{zz}$) & ( 0.000,  0.000, 0.362) &   0 & 0 & 0.93669 \\  
                 \hline
                                            &  $G'$ & \textrm{RT-CCSD(dp)} & R$^{2}_{x}$(dp) & R$^{2}_{y}$(dp)& R$^{2}_{z}$(dp)\\
                                          \hline
                \textrm{QRCW} & ($G'_{xx}$, $G'_{yx}$, $G'_{zx}$) & (-0.387, -0.097, 0.000) &   0.99956 & 0.99956 & 0 \\
                                          & ($G'_{xy}$, $G'_{yy}$, $G'_{zy}$) & ( 0.058,  0.013, 0.000) &   0.99915 & 0.99896 & 0 \\
                                          & ($G'_{xz}$, $G'_{yz}$, $G'_{zz}$) & ( 0.000,  0.000, 0.362) &   0 & 0 & 0.99597 \\  
                 \hline
                                            &  $G'$ & \textrm{RT-CCSD(sp)} & R$^{2}_{x}$(sp) & R$^{2}_{y}$(sp)& R$^{2}_{z}$(sp)\\
                                          \hline
                \textrm{QRCW} & ($G'_{xx}$, $G'_{yx}$, $G'_{zx}$) & (-0.387, -0.097, 0.000) &   0.99956 & 0.99956 & 0 \\
                                          & ($G'_{xy}$, $G'_{yy}$, $G'_{zy}$) & ( 0.058,  0.013, 0.000) &   0.99915 & 0.99896 & 0 \\
                                          & ($G'_{xz}$, $G'_{yz}$, $G'_{zz}$) & ( 0.000,  0.000, 0.362) &   0 & 0  & 0 .99597\\  
                
   \end{tabular}
  \label{tab:g-ccsd}
\end{table}
% Table9-G' tensor
\begin{table}
  \centering
    \caption{\ZW{RT-CC3/cc-pVDZ} $G'$ tensor elements (in atomic units) of H$_{2}$
    dimer at 582 nm obtained using single- and double-precision computations,
    with linear ramped continuous wave (LRCW) and quadratic ramped continuous
    wave (QRCW) fields. The R$^{2}$ values, indicating the quality of curve
    fitting, are displayed in the last three columns.}
  \begin{tabular}{c|c|c|ccc}
                                          &  $G'$ & \textrm{RT-CC3(dp)} & R$^{2}_{x}$(dp) & R$^{2}_{y}$(dp)& R$^{2}_{z}$(dp)\\
                                          \hline
                \textrm{LRCW} & ($G'_{xx}$, $G'_{yx}$, $G'_{zx}$) & (\ZW{-0.387}, -0.097, 0.000) &   \ZW{0.93654} & \ZW{0.93842} & 0 \\
                                          & ($G'_{xy}$, $G'_{yy}$, $G'_{zy}$) & ( 0.058,  0.013, 0.000) &   \ZW{0.87831} & \ZW{0.85893} & 0\\
                                          & ($G'_{xz}$, $G'_{yz}$, $G'_{zz}$) & ( 0.000,  0.000, \ZW{0.362}) &   0 & 0 & \ZW{0.93569} \\   
                \hline
                                           &  $G'$ & \textrm{RT-CC3(sp)} & R$^{2}_{x}$(sp) & R$^{2}_{y}$(sp)& R$^{2}_{z}$(sp)\\
                                          \hline
                \textrm{LRCW} & ($G'_{xx}$, $G'_{yx}$, $G'_{zx}$) & (\ZW{-0.387}, -0.097, 0.000) &   0.93654 & 0.93842 & 0 \\
                                          & ($G'_{xy}$, $G'_{yy}$, $G'_{zy}$) & ( 0.058,  0.013, 0.000) &   0.87831 & 0.85893 & 0 \\
                                          & ($G'_{xz}$, $G'_{yz}$, $G'_{zz}$) & ( 0.000,  0.000, \ZW{0.362}) &   0 & 0 & 0.93569 \\  
                 \hline
                                            &  $G'$ & \textrm{RT-CC3(dp)} & R$^{2}_{x}$(dp) & R$^{2}_{y}$(dp)& R$^{2}_{z}$(dp)\\
                                          \hline
                \textrm{QRCW} & ($G'_{xx}$, $G'_{yx}$, $G'_{zx}$) & (-0.387, -0.097, 0.000) &   0.99955 & 0.99955 & 0 \\
                                          & ($G'_{xy}$, $G'_{yy}$, $G'_{zy}$) & ( 0.058,  0.013, 0.000) &   0.99911 & 0.99891 & 0 \\
                                          & ($G'_{xz}$, $G'_{yz}$, $G'_{zz}$) & ( 0.000,  0.000, 0.362) &   0 & 0 & 0.99602 \\  
                 \hline
                                            &  $G'$ & \textrm{RT-CC3(sp)} & R$^{2}_{x}$(sp) & R$^{2}_{y}$(sp)& R$^{2}_{z}$(sp)\\
                                          \hline
                \textrm{QRCW} & ($G'_{xx}$, $G'_{yx}$, $G'_{zx}$) & (-0.387, -0.097, 0.000) &   0.99955 & 0.99955 & 0 \\
                                          & ($G'_{xy}$, $G'_{yy}$, $G'_{zy}$) & ( 0.058,  0.013, 0.000) &  0.99911 & 0.99891 & 0 \\
                                          & ($G'_{xz}$, $G'_{yz}$, $G'_{zz}$) & ( 0.000,  0.000, 0.362) &   0 & 0 & \ZW{0.99601} \\  
                
   \end{tabular}
  \label{tab:g-cc3} 
\end{table}

With the same data set, it is also possible to extract
quadratic response functions of the form $\langle\!\langle
\boldsymbol{\hat{m}};\boldsymbol{\hat{\mu}},\boldsymbol{\hat{\mu}}\rangle\!\rangle_{\omega,\omega^\prime}$
using the second-order induced magnetic dipole moments.
Such response functions describe magnetic-dipole second harmonic generation
for $\omega^\prime = \omega$, while for $\omega^\prime = -\omega$ they are related
to Verdet's constant and magnetic optical rotation.~\cite{Parkinson1997,Coriani1997}
The latter can, of course, also be obtained directly from the polarizability in a
finite magnetic field.~\cite{Ofstad2023_magnetic}
As shown in Fig.~\ref{fig:g-fit},
$\langle\!\langle
\hat{m}_z;\hat{\mu}_z,\hat{\mu}_z\rangle\!\rangle_{\omega,\omega}$ and
$\langle\!\langle
\hat{m}_z;\hat{\mu}_z,\hat{\mu}_z\rangle\!\rangle_{\omega,-\omega}$ can be
obtained as the amplitude and phase of the fitted curve, respectively, and 
RT-CCSD and RT-CC3 results are listed in Table~\ref{tab:mag-resp-ftn}.
No significant difference is observed between single- and double-precision results for both of
the response functions, although the $R^2$ value of the single-precision
calculation is slightly lower. It is obvious that the quality of the curve fitting
is not sufficient to provide robust results, especially for the LRCW cases.
Compared to the hyperpolarizability results obtained from
$\mu_{\alpha\beta\beta}^{(2)}$ in Table~\ref{tab:hyp-shg} and
~\ref{tab:hyp-or}, the second-order response of the magnetic dipole moments to
the external electric field is much weaker. The results are more sensitive to
the choice of the external field, the length of the propagation, the numerical
differentiation, etc. \ZW{For example, the relative difference between the LRCW and
the QRCW results of RT-CC3 (dp)
$\langle\!\langle\hat{m}_z;\hat{\mu}_z,\hat{\mu}_z\rangle\!\rangle_{\omega,\omega}$
(9.15\%) is larger than the relative difference of $\beta_{zzz}$
(SHG) (0.49\%).} QRCW with extra ramped cycles is assumed to be necessary and
helpful to improve the quality of the curve fitting and provide reliable results.
In that case, the RT-CC framework would be practical for calculating
$\langle\!\langle\hat{m}_z;\hat{\mu}_z,\hat{\mu}_z\rangle\!\rangle$
straightforwardly.

% Table10-<<m;mu,mu>>
\begin{table}
  \centering
  \caption{RT-CCSD/cc-pVDZ and RT-CC3/cc-pVDZ $\langle\!\langle \hat{m}_z;\hat{\mu}_z,\hat{\mu}_z\rangle\!\rangle$
  (in atomic units) of \ch{H_2} dimer at 582 nm obtained from simulations with
  linear ramped continuous wave (LRCW) and quadratic ramped continuous wave
  (QRCW) fields. The R$^{2}$ values, reflecting the quality of curve fitting,
  are displayed in the last column.}
  \begin{tabular}{c|c|cc|c}
    &  \textrm{Method}  & $\langle\!\langle \hat{m}_z;\hat{\mu}_z,\hat{\mu}_z\rangle\!\rangle_{\omega,\omega}$ 
                        & $\langle\!\langle \hat{m}_z;\hat{\mu}_z,\hat{\mu}_z\rangle\!\rangle_{\omega,-\omega}$ 
                        & R$^{2}$ \\
    \hline                   
     \textrm{LRCW} & \textrm{RT-CCSD\ (dp)} & -0.147 & -0.001 & 0.51509 \\
                   & \textrm{RT-CCSD\ (sp)} & -0.147 & -0.001 & 0.47038 \\
    \hline
     \textrm{QRCW} & \textrm{RT-CCSD\ (dp)} & -0.161 & -0.001 & 0.91640 \\
                   & \textrm{RT-CCSD\ (sp)} & -0.161 & -0.001 & 0.89186 \\
    \hline\hline
    \textrm{\ZW{LRCW}} & \textrm{RT-CC3\ (dp)} & -0.149 &  -0.002 & 0.55592 \\
                  & \textrm{RT-CC3\ (sp)} & -0.149 &  -0.002 & 0.54161 \\
      \hline
    \textrm{\ZW{QRCW}} & \textrm{RT-CC3\ (dp)} & -0.164 & -0.001 & 0.91596 \\
                  & \textrm{RT-CC3\ (sp)} & -0.163 & -0.002 & 0.87705 \\
  \end{tabular}
  \label{tab:mag-resp-ftn}
\end{table}

%% file: table_rtcc3_vs_tdnoccd.tex
% Table7-RTCC3-vs-TDNOCCD
\begin{table}
\centering
\caption{Polarizabilities (in atomic units) of \ch{Ne}, \ch{HF}, \ch{H2O},
\ch{NH3} and \ch{CH4}.}
\begin{tabular}{l l r r r r r r r r r}
\hline
\hline
\ch{Ne}&$\omega\,(\text{a.u.})$ & $0.1$ & $0.2$ & $0.3$ & $0.4$ & $0.5$ \\
\hline
&FCI        &   $2.70$ & $2.79$ & $2.97$ & $3.31$ & $4.09$  \\
&LR-CC3      &   $2.71$ & $2.80$ & $2.98$ & $3.32$ & $4.10$  \\
&\ZW{RT-CC3}      &   $2.71$ & $2.80$ & $2.99$ & $3.67$ & $4.11$  \\
&LR-CCSD     &   $2.74$ & $2.83$ & $3.01$ & $3.38$ & $4.23$  \\
&RT-CCSD     &   $2.74$ & $2.83$ & $3.03$ & $3.49$ & $4.76$  \\
&TDNOCCD    &   $2.66$ & $2.75$ & $2.92$ & $3.41$ & $3.93$   \\
\hline
\ch{HF}&$\omega\,(\text{a.u.})$        & \multicolumn{2}{c}{$0.1$}&  \multicolumn{2}{c}{$0.2$} & \multicolumn{2}{c}{$0.3$}            \\
&                      & $\alpha_{yy}$   & $\alpha_{zz}$ & $\alpha_{yy}$   & $\alpha_{zz}$ & $\alpha_{yy}$   & $\alpha_{zz}$ \\
\hline
&FCI      & $4.39$ & $6.33$ & $4.76$ & $6.74$ & $6.00$ & $7.63$ \\
&LR-CC3    & $4.39$ & $6.34$ & $4.77$ & $6.76$ & $6.03$ & $7.64$ \\
&\ZW{RT-CC3}    & $4.39$ & $6.34$ & $\ZW{4.73}$ & $\ZW{6.75}$ & $6.52$ & $\ZW{7.48}$ \\
&LR-CCSD   &     $4.44$      &   $6.41$  &     $4.83$      &  $6.83$ &     $6.19$      &   $7.73$          \\
&RT-CCSD   & $4.45$ & $6.41$ & $4.84$ & $6.83$ & $6.72$ & $7.84$\\
&TDNOCCD  & $4.30$ & $6.21$ & $4.63$ & $6.60$ & $6.09$ & $7.35$ \\
\hline
\ch{H2O} & $\omega\,(\text{a.u.})$ & \multicolumn{3}{c}{$0.0428$} & \multicolumn{3}{c}{$0.0656$} & \multicolumn{3}{c}{$0.1$} \\
&                      & $\alpha_{xx}$          & $\alpha_{yy}$   & $\alpha_{zz}$ & $\alpha_{xx}$  & $\alpha_{yy}$   & $\alpha_{zz}$ & $\alpha_{xx}$  & $\alpha_{yy}$   & $\alpha_{zz}$ \\
\hline
&LR-CC3   &     $8.72$    &     $9.86$       &   $9.04$      &     $8.83$    &     $9.92$       &   $9.12$      &     $9.10$    &     $10.06$      &   $9.30$\\
&RT-CC3   &     $8.72$    &     $9.86$       &   $9.04$      &     $8.83$    &     $9.92$      &   $9.12$      &     $9.10$    &     $10.06$      &   $9.31$\\
&LR-CCSD   &     $8.78$    &     $9.93$       &   $9.11$      &     $8.89$    &     $9.99$       &   $9.19$      &     $9.18$    &     $10.14$      &   $9.37$\\
&RT-CCSD   &     $8.78$    &     $9.93$       &   $9.11$      &     $8.90$    &     $10.00$      &   $9.19$      &     $9.19$    &     $10.14$      &   $9.37$\\
&TDNOCCD  &     $8.45$       &   $9.67$               & $8.84$              &  $8.55$             &  $9.73$                &   $8.90$            &     $8.79$    &     $9.86$       &   $9.07$\\
\hline
\ch{NH3} & $\omega\,(\text{a.u.})$        & \multicolumn{2}{c}{$0.0428$} & \multicolumn{2}{c}{$0.0656$} & \multicolumn{2}{c}{$0.1$} \\
&        & $\alpha_{yy}$   & $\alpha_{zz}$ & $\alpha_{yy}$& $\alpha_{zz}$ & $\alpha_{yy}$& $\alpha_{zz}$ \\
\hline
&LR-CC3   &      $13.05$    &    $15.02$    & $13.15$      &   $15.33$     & $13.39$      & $16.14$       \\
&RT-CC3   &      $13.05$    &    $15.02$    & $13.15$      &   $15.33$     & $13.38$      & $\ZW{15.99}$       \\
&LR-CCSD   &      $13.10$    &    $15.04$    & $13.20$      &   $15.35$     & $13.44$      & $16.15$       \\
&RT-CCSD   &      $13.10$    &    $15.05$    & $13.20$      &   $15.36$     & $13.45$      & $16.15$       \\
&TDNOCCD  &      $12.85$    &    $14.57$    & $12.94$      &   $14.84$     & $13.17$      & $15.51$       \\
\hline
\ch{CH4}&$\omega\,(\text{a.u.})$ & $0.0656$ & $0.1$ & $0.2$\\
\hline
&LR-CC3   & $17.04$ & $17.38$ & $19.55$ \\
&RT-CC3   & $17.04$ & $17.37$ & $\ZW{19.34}$ \\
&LR-CCSD   & $17.05$ & $17.39$ & $19.55$ \\
&RT-CCSD   & $17.05$ & $17.39$ & $19.58$ \\
&TDNOCCD  & $16.86$ & $17.19$ & $19.22$ \\
\hline
\hline
\end{tabular}
\label{tab:noccd}
\end{table}

%% file: conc.tex
\section{Conclusions}\label{conc_cc3}
The RT-CC3 method has been implemented with additional single-precision and GPU
options. The working equations of RT-CC3 in the closed-shell and spin-adapted
formalism are provided, with considerations of optimizing the performance in
terms of reducing the number of higher-order tensor contractions. The
implementation has been validated through the calculation of the absorption
spectrum of \ch{H_{2}O} in both single- and double-precision. Numerical
experiments have also been conducted with water clusters to test the
computational cost of RT-CC3 simulations. It has been found that the use of
GPUs can significantly speed up calculations by up to a factor of 17 due to the
computational power they provide for tensor contractions. The acceleration
gained from utilizing either GPUs or single-precision arithmetic needs to be
observed significantly within a relatively large system (e.g., 72 molecular
orbitals). To achieve the theoretical speedup, a much larger system is needed,
however, optimization of memory allocation will also need to be taken into
account because of the limited memory available on GPUs and the overhead of
data migration. With the promising results of our Python implementation,
exploring a productive-level code is worthwhile, especially for making the
RT-CC3 method, which scales as ${\cal O}(N^{7})$, feasible for large system/basis set
and/or long RT propagations.  

For the calculation of optical properties, we have demonstrated that RT-CC3 is
a feasible tool to obtain dynamic polariazabilities, first
hyperpolarizabilities, and the $G'$ tensor with good agreement to LR-CC3 and a
reasonable computational cost. The type of applied field, the precision
arithmetic, and the level of theory were tested with \ch{H_{2}O} and \ch{H_{2}} dimer. It has been
demonstrated through all test cases, including our new RT-CC3 method, that the QRCW
can substantially improve curve fitting and requires only two
optical cycles for propagation. 
Especially for first hyperpolarizabilites, the
curve of the second-order dipole moments from LRCW calculation has
`discontinuities' in some places, leading to a large error and a low R$^{2}$
value for curve fitting. The QRCW is required here to obtain reliable results.
The same is found in $G'$ tensor results, where some shifts appear in the curve
of the induced magnetic dipole moments from LRCW calculations but not the QRCW
ones. Regarding the single-precision calculations, no discrepancy is found in
the polarizabilities and $G'$ tensor elements that are associated with the
first derivative of electric and magnetic dipole moments, respectively. A
significant difference, however, is found in the first hyperpolarizabilities.
Although the QRCW can still reduce the error compared to the LRCW,
single-precision results remain less accurate compared to double-precision
results. With the same set of induced magnetic dipole moments for obtaining $G'$ tensor, we can also
extract the quadratic response function 
$\langle\!\langle\boldsymbol{\hat{m}};\boldsymbol{\hat{\mu}},\boldsymbol{\hat{\mu}}\rangle\!\rangle_{\omega,\omega^\prime}$,
although QRCW with extra ramped cycles is assumed to be needed for more accurate results. 
These conclusions hold for both RT-CCSD and RT-CC3. 

Additionally, ten-electron systems including \ch{Ne}, \ch{HF}, \ch{H_{2}O},
\ch{NH_{3}} and \ch{CH_{4}} are used to test the calculation of
polarizabilities with RT-CCSD, RT-CC3, and particularly TDNOCCD. It has been
observed that the accuracy drops significantly when the frequency is closer to
the resonance, while for the small frequencies, RT-CC3 matches LR-CC3 and FCI
with errors less than 0.1\%. The trend of the error of RT-CC3 is consistent
with RT-CCSD for most cases, except for the two values with the highest chosen
frequencies of \ch{NH_{3}} and \ch{CH_{4}}. We have shown that the CC3
polarizabilities increase slightly faster when the frequency moves towards
resonance, which may lead to a larger error. The TDNOCCD results show that the
explicit orbital optimization lowers the polarizability values compared to
RT-CCSD, where the only difference in these two methods is the orbital
optimization. Compared to RT-CC3, TDNOCCD results are closer to LR-CC3/FCI
results when RT-CC3 largely overestimates the results, otherwise, RT-CC3 yields
more accurate results. 

%% file: extra.tex
%extra.tex%

\section{Supporting Information} \label{si}
Atomic coordinates of all molecular test cases.

\section{Acknowledgements} \label{ack} 
This work was supported by the U.S.\ National Science Foundation via
grant CHE-2154753 and by the Research Council of Norway through its Centres of
Excellence Scheme, Grant. No. 262695.
The authors are grateful to Advanced Research Computing at Virginia Tech for providing
computational resources that have contributed to the
results reported within the paper.

%% file: appendix.tex
\section{Appendix: Derivation of the RT-CC3 Equations}
\renewcommand{\theequation}{A\arabic{equation}}
\setcounter{equation}{0}

Assuming a closed-shell spin-restricted reference determinant, the Hamiltonian can be written in the canonical basis as
\begin{align}
  &\hat{H}(t) = \hat{F} + \hat{U} + \beta \hat{V}(t), \\
  &\hat{F} = \sum_p \epsilon_p \hat{E}_{pp}, \\
  &\hat{U} = \hat{H}(t) - \beta \hat{V}(t) - \hat{F}.
\end{align}
The $T_1$-transformed Hamiltonian is (dropping the reference to time dependence for convenience):
\begin{equation}
  H = \mathrm{e}^{-\hat{T}_1} \hat{H} \mathrm{e}^{\hat{T}_1} = F + U + \beta V,
\end{equation}
where $F = \exp(-\hat{T}_1) \hat{F} \exp(\hat{T}_1)$ and analogously for $U$ and $V$.
We assume that $\hat{V}$ is a one-electron operator. $\beta$ is the field strength.

The time-dependent CCSDT Lagrangian can be written as
\begin{align}
  \mathcal{L}_\text{CCSDT} &= \lambda_0 \braket{\Phi_0\vert H + [H,\hat{T}_2] \vert \Phi_0} \nonumber \\
  &+ \braket{\Phi_0 \vert \hat{\Lambda}_1 \left( H + [H,\hat{T}_2] + [H,\hat{T}_3] \right) \vert \Phi_0} \nonumber \\
  &+ \braket{\Phi_0 \vert \hat{\Lambda}_2 \left( H + [H,\hat{T}_2] + [H,\hat{T}_3] + \frac{1}{2} [[H,\hat{T}_2],\hat{T}_2] \right) \vert \Phi_0}
  \nonumber \\
  &+ \braket{\Phi_0 \vert \hat{\Lambda}_3 \left( [H,\hat{T}_2] + [H,\hat{T}_3] + \frac{1}{2} [[H,\hat{T}_2],\hat{T}_2] 
                                               + [[H,\hat{T}_2],\hat{T}_3] \right) \vert \Phi_0}
  \nonumber \\
  &- \mathrm{i} \lambda_0 \dot{t}_0 - \mathrm{i} \sum_{\mu_1} \lambda_{\mu_1} \dot{t}{\mu_1}
  - \mathrm{i} \sum_{\mu_2} \lambda_{\mu_2} \dot{t}_{\mu_2} - \mathrm{i} \sum_{\mu_3} \lambda_{\mu_3} \dot{t}_{\mu_3},
\end{align}
To obtain the CC3 Lagrangian we now apply the following rules regarding order in the fluctuation potential $\hat{U}$,
\begin{alignat*}{2}
    &\text{Order}\ 0: &\qquad & \hat{F}, \hat{V}, \hat{T}_1, \hat{\Lambda}_1 \\
    &\text{Order}\ 1: &\qquad & \hat{U}, \hat{T}_2, \hat{\Lambda}_2 \\
    &\text{Order}\ 2: &\qquad & \hat{T}_3, \hat{\Lambda}_3
\end{alignat*}
and \emph{neglect all terms above fourth order in the CCSDT Lagrangian}:
\begin{align}
  \mathcal{L}_\text{CC3} &= \lambda_0 \braket{\Phi_0\vert H + [H,\hat{T}_2] \vert \Phi_0} \nonumber \\
  &+ \braket{\Phi_0 \vert \hat{\Lambda}_1 \left( H + [H,\hat{T}_2] + [H,\hat{T}_3] \right) \vert \Phi_0} \nonumber \\
  &+ \braket{\Phi_0 \vert \hat{\Lambda}_2 \left( H + [H,\hat{T}_2] + [H,\hat{T}_3] + \frac{1}{2} [[H,\hat{T}_2],\hat{T}_2] \right) \vert \Phi_0}
  \nonumber \\
  &+ \braket{\Phi_0 \vert \hat{\Lambda}_3 \left( [H,\hat{T}_2] + [F+\beta V,\hat{T}_3] + \frac{1}{2} [[\beta V,\hat{T}_2],\hat{T}_2] \right) \vert \Phi_0}
  \nonumber \\
  &- \mathrm{i} \lambda_0 \dot{t}_0 - \mathrm{i} \sum_{\mu_1} \lambda_{\mu_1} \dot{t}_{\mu_1}
  - \mathrm{i} \sum_{\mu_2} \lambda_{\mu_2} \dot{t}_{\mu_2} - \mathrm{i} \sum_{\mu_3} \lambda_{\mu_3} \dot{t}_{\mu_3}.
\end{align}
The EOMs can now be obtained from the Euler-Lagrange equations
\begin{equation}
    \frac{\partial \mathcal{L}_\text{CC3}}{\partial z} - \frac{\mathrm{d}}{\mathrm{d} t} \frac{\partial \mathcal{L}_\text{CC3}}{\partial \dot{z}} = 0,
\end{equation}
where $z$ denotes the cluster amplitudes $t$ and $\lambda$.

Setting $z = \lambda_0$, we find
\begin{equation}
    \mathrm{i}\dot{t}_0 = \braket{\Phi_0\vert H + [H,\hat{T}_2] \vert \Phi_0}.
\end{equation}
Setting $z = t_0$, we find
\begin{equation}
    \dot{\lambda}_0 = 0,
\end{equation}
which means that we may choose $\lambda_0 = 1$.

The remaining $\hat{T}$ equations read (obtained by setting $z=\lambda_{\mu_i},\ i=1,2,3$):
\begin{align}
    \mathrm{i} \dot{t}_{\mu_1} &=
    \braket{\mu_1 \vert H + [H,\hat{T}_2] + [H,\hat{T}_3] \vert \Phi_0}, \\
    \mathrm{i} \dot{t}_{\mu_2} &=
    \braket{\mu_2 \vert H + [H,\hat{T}_2] + [H,\hat{T}_3] + \frac{1}{2} [[H,\hat{T}_2],\hat{T}_2] \vert \Phi_0}, \\
    \mathrm{i} \dot{t}_{\mu_3} &=
    \braket{\mu_3 \vert [H,\hat{T}_2] + [F+\beta V,\hat{T}_3] + \frac{1}{2} [[\beta V,\hat{T}_2],\hat{T}_2] \vert \Phi_0},
\end{align}
and the $\hat{\Lambda}$ equations read (obtained by setting $z = t_{\mu_i},\ i=1,2,3$):
\begin{align}
    -\mathrm{i} \dot{\lambda}_{\mu_1} &= \braket{\Phi_0\vert [H, \tau_{\mu_1}] \vert \Phi_0} \nonumber \\
    &+ \braket{\Phi_0 \vert \hat{\Lambda}_1 \left( [H, \tau_{\mu_1}] + [[H, \tau_{\mu_1}],\hat{T}_2] \right) \vert \Phi_0} \nonumber \\
    &+ \braket{\Phi_0 \vert \hat{\Lambda}_2 \left( [H, \tau_{\mu_1}] + [[H, \tau_{\mu_1}],\hat{T}_2] 
                                                 + [[H, \tau_{\mu_1}],\hat{T}_3] \right) \vert \Phi_0} \nonumber \\
    &+ \braket{\Phi_0 \vert \hat{\Lambda}_3 \left( [H, \tau_{\mu_1}],\hat{T}_2] + [[\beta V, \tau_{\mu_1}],\hat{T}_3] \right) \vert \Phi_0},\\
    -\mathrm{i} \dot{\lambda}_{\mu_2} &=  \braket{\Phi_0\vert [H,\tau_{\mu_2}] \vert \Phi_0} \nonumber \\
    &+ \braket{\Phi_0 \vert \hat{\Lambda}_1 [H, \tau_{\mu_2}] \vert \Phi_0} \nonumber \\
    &+ \braket{\Phi_0 \vert \hat{\Lambda}_2 \left( [H, \tau_{\mu_2}] + [[H, \tau_{\mu_2}],\hat{T}_2] \right) \vert \Phi_0}
  \nonumber \\
    &+ \braket{\Phi_0 \vert \hat{\Lambda}_3 \left( [H, \tau_{\mu_2}] + [[\beta V, \tau_{\mu_2}],\hat{T}_2] \right) \vert \Phi_0}, \\
    -\mathrm{i} \dot{\lambda}_{\mu_3} &=
    \braket{\Phi_0 \vert \hat{\Lambda}_1 [H, \tau_{\mu_3}] \vert \Phi_0} \nonumber \\
    &+ \braket{\Phi_0 \vert \hat{\Lambda}_2 [H, \tau_{\mu_3}] \vert \Phi_0}
  \nonumber \\
    &+ \braket{\Phi_0 \vert \hat{\Lambda}_3 [F+\beta V, \tau_{\mu_3}] \vert \Phi_0}.
\end{align}
Note that
\begin{equation}
    [F, \hat{T}_3] = [\hat{F}, \hat{T}_3] = \sum_{\mu_3} \epsilon_{\mu_3} t_{\mu_3} \tau_{\mu_3}
\end{equation}

The right-hand sides of the singles and doubles equations can be separated into CCSD components 
and triples specific terms. To derive the spin-adapted 
expression of the amplitude equations, we write the singles, doubles, and triples cluster operators as
\begin{equation}
\hat{T}_{1}=\sum_{ia}t_{i}^{a}E_{ai},
\label{eq:cc3-T1}
\end{equation}
\begin{equation}
\hat{T}_{2}=\frac{1}{2}\sum_{ijab}t_{ij}^{ab}E_{ai}E_{bj},
\label{eq:cc3-T2}
\end{equation}
and
\begin{equation}
\hat{T}_{3}=\frac{1}{6}\sum_{ijkabc}t_{ijk}^{abc}E_{ai}E_{bj}E_{ck},
\label{eq:cc3-T3}
\end{equation}
respectively, where $i, j, k, ...$ are occupied orbitals, $a, b, c, ...$ are
virtual orbitals, and the unitary group generators are defined as
\begin{equation}
\label{eq:unitary-group-gen}
E_{pq}=a^{+}_{p_{\alpha}}a_{q_{\alpha}}+a^{+}_{p_{\beta}}a_{q_{\beta}}
\end{equation}
with $p, q$ being molecular orbitals. By inserting this form of $\hat{T}$ amplitudes 
and the similar form for the $\hat{\Lambda}$ amplitudes into the triples specific terms, 
the spin-adapted expression can be written as
\begin{equation}
X_{i}^{a}= \sum_{jkbc}(t_{ijk}^{abc}-t_{ijk}^{cba})L_{jkbc}
\label{eq:cc3-x1-exp}
\end{equation}
in the $\hat{T}_1$ equation,
\begin{align}
\begin{split}
X_{ij}^{ab}&=P_{ij}^{ab}\left\{\sum_{kc}(t_{ijk}^{abc}-t_{ijk}^{cba})\tilde{f}_{kc} +
\sum_{kce}(2t_{ijk}^{cbe}-t_{ijk}^{ceb}-t_{ijk}^{ebc}) \tilde{\langle ak|ce \rangle} \right. \\
 &\left. - \sum_{kmc}(2t_{mjk}^{cba}-t_{mjk}^{bca}-t_{mjk}^{abc}) \tilde{\langle km|ic \rangle} \right\}
\label{eq:cc3-x2-exp}
\end{split}
\end{align}
in the $\hat{T}_2$ equation,
\begin{align}
\begin{split}
Y_{a}^{i} &= \sum\limits_{\substack{jkl\\bcd}}\Bigl[(t_{jkl}^{bcd}\langle ij|cd \rangle \lambda_{ab}^{kl} + t_{jkl}^{bcd}
\langle kl|ab \rangle \lambda_{cd}^{ij}+ t_{jkl}^{bcd}L_{ij}^{ab}\lambda_{cd}^{kl}) \\
 &-(t_{jkl}^{bcd}L_{ijac}\lambda_{bd}^{kl} + t_{jkl}^{bcd}L_{jibc}\lambda_{ad}^{kl} + t_{jkl}^{bcd}L_{ijab}\lambda_{cd}^{il})\\
 &\left. -(t_{jk}^{bc} \tilde{\langle jd|la \rangle} \lambda_{bcd}^{lki} + t_{jk}^{bc} \tilde{\langle dj|la \rangle} \lambda_{bcd}^{ikl} 
   + t_{jk}^{bc} \tilde{\langle id|lb \rangle} \lambda_{acd}^{lkj} + t_{jk}^{bc} \tilde{\langle di|la \rangle}
   \lambda_{acd}^{jkl})\right] \\
 &+\sum\limits_{\substack{jk\\bcde}} t_{jk}^{bc} \tilde{\langle de|ab \rangle} \lambda_{cde}^{kij}
   +\sum\limits_{\substack{jklm\\bc}}t_{jk}^{bc} \tilde{\langle ij|lm \rangle} \lambda_{abc}^{lmk}
\label{eq:cc3-y1-exp}
\end{split}
\end{align}
in the $\hat{\Lambda}_1$ equation, and
\begin{equation}
Y_{ab}^{ij}= P_{ij}^{ab} \left\{ \sum_{lde} \tilde{\langle de|al \rangle} \lambda_{dbe}^{ijl} -  \sum_{lmd} \tilde{\langle
id|ml \rangle} \lambda_{abd}^{mjl} \right\}
\label{eq:cc3-y2-exp}
\end{equation}
in the $\hat{\Lambda}_2$ equation. In the time-independent case, the triples can be calculated as
\begin{align}
\begin{split}
t_{ijk}^{abc} &= -{\epsilon_{ijk}^{abc}}^{-1} P_{ijk}^{abc} \left\{ \sum_{e}t_{ij}^{ae} 
                 \tilde{\langle cb|ke \rangle} - \sum_{m}t_{im}^{ab} 
		 \tilde{\langle mc|jk \rangle} \right. \\
              &\left. -\frac{\beta}{2}( \sum_d v_{da}t_{ijk}^{dbc}+ \sum_l v_{il}t_{ljk}^{abc}) 
           - \beta \sum_{ld} v_{ld}t_{ij}^{ad}t_{kl}^{cb} \right\}
\end{split}
\end{align}
and
\begin{align}
\begin{split}
\lambda_{abc}^{ijk}&=P_{ijk}^{abc}\left\{ (L_{ijab}\lambda_{c}^{k} - L_{ijac}\lambda_{b}^{k}) 
  + (\tilde{f}_{ia}\lambda_{bc}^{jk} + \sum_{l} \tilde{\langle kj|al \rangle} \lambda_{bc}^{li} - \sum_{d} \tilde{\langle
  kd|ab \rangle} \lambda_{cd}^{ij}) \right. \\
 &\left. + \frac{1}{2}P_{ij}^{ab}(-\tilde{f}_{ja}\lambda_{bc}^{ik} - \sum_{l} \tilde{L}_{ijal}\lambda_{bc}^{lk} + \sum_{d} \tilde{L}_{djab}\lambda_{cd}^{ki})
  + \frac{\beta}{2} ( \sum_d v_{da}\lambda^{ijk}_{dbc} - \sum_l v_{il}\lambda^{ljk}_{abc})\right\},
\label{eq:cc3-l3-exp}
\end{split}
\end{align}
where $v_{pq}$ is the matrix element of the perturbation operator $\hat{V}$.  We emphasize that, in the perturbed case
(i.e., non-zero $\beta V$), the triples cannot be computed via a closed expression in terms of the $\hat{T}_1$ and
$\hat{T}_2$ amplitudes, but must be computed iteratively.

The one- and two-electron integrals, $f_{pq}$ and $\langle pq|rs \rangle$, are extracted from the one- and two-electron component of the Hamiltonian, respectively, as 
\begin{equation}
\hat{H} = \sum_{pq}f_{pq}\{E_{pq}\} + \frac{1}{2}\sum_{pqrs}\langle pq|rs \rangle \{E_{pq}E_{rs}\},
\end{equation}
where $\{E_{pq}\}$ denotes the normal ordering unitary-group generator. $L_{pqrs}$ is defined as
\begin{equation}
L_{pqrs} = 2\langle pq|rs \rangle - \langle pq|sr \rangle.
\end{equation}
$\tilde{f}_{pq}$, $\tilde{\langle pq|rs \rangle}$ and $\tilde{L}_{pqrs}$ are components of the $T_{1}$-transformed Hamiltonian. The permutation operators are defined as 
\begin{equation}
P_{ij}^{ab}f_{ij}^{ab} = f_{ij}^{ab} + f_{ji}^{ba}
\label{eq:cc3-pijab}
\end{equation}
and
\begin{equation}
P_{ijk}^{abc}f_{ijk}^{abc} = f_{ijk}^{abc} + f_{jik}^{bac} + f_{ikj}^{acb} + f_{kji}^{cba} + f_{kij}^{cab} + f_{jki}^{bca}.
\label{eq:cc3-pijkabc}
\end{equation}
The explicit formula of the additional terms involving triples in the one-electron density can be written as
\begin{equation}
D_{ij} = -\frac{1}{2}\sum\limits_{\substack{kl\\abc}}t_{ilk}^{abc}\lambda_{abc}^{jlk},
\label{eq:cc3-dij}
\end{equation}
\begin{equation}
D_{ab} = \frac{1}{2}\sum\limits_{\substack{ijk\\cd}}t_{ijk}^{bdc}\lambda_{adc}^{ijk},
\end{equation}
and
\begin{equation}
D_{ia} = \sum_{jkbc}(t_{ijk}^{abc} - t_{ijk}^{bac})\lambda_{bc}^{jk} - \sum\limits_{\substack{jkl\\bcd}}\lambda_{bcd}^{jkl}t_{il}^{cd}t_{kj}^{ab}.
\end{equation}

%% file: main.bbl
\providecommand{\latin}[1]{#1}
\makeatletter
\providecommand{\doi}
  {\begingroup\let\do\@makeother\dospecials
  \catcode`\{=1 \catcode`\}=2 \doi@aux}
\providecommand{\doi@aux}[1]{\endgroup\texttt{#1}}
\makeatother
\providecommand*\mcitethebibliography{\thebibliography}
\csname @ifundefined\endcsname{endmcitethebibliography}
  {\let\endmcitethebibliography\endthebibliography}{}
\begin{mcitethebibliography}{73}
\providecommand*\natexlab[1]{#1}
\providecommand*\mciteSetBstSublistMode[1]{}
\providecommand*\mciteSetBstMaxWidthForm[2]{}
\providecommand*\mciteBstWouldAddEndPuncttrue
  {\def\EndOfBibitem{\unskip.}}
\providecommand*\mciteBstWouldAddEndPunctfalse
  {\let\EndOfBibitem\relax}
\providecommand*\mciteSetBstMidEndSepPunct[3]{}
\providecommand*\mciteSetBstSublistLabelBeginEnd[3]{}
\providecommand*\EndOfBibitem{}
\mciteSetBstSublistMode{f}
\mciteSetBstMaxWidthForm{subitem}{(\alph{mcitesubitemcount})}
\mciteSetBstSublistLabelBeginEnd
  {\mcitemaxwidthsubitemform\space}
  {\relax}
  {\relax}

\bibitem[Gauss(1998)]{Gauss98}
Gauss,~J. In \emph{Encyclopedia of Computational Chemistry}; Schleyer,~P.,
  Allinger,~N.~L., Clark,~T., Gasteiger,~J., Kollman,~P.~A., {Schaefer
  III},~H.~F., Schreiner,~P.~R., Eds.; John Wiley and Sons: Chichester, 1998;
  pp 615--636\relax
\mciteBstWouldAddEndPuncttrue
\mciteSetBstMidEndSepPunct{\mcitedefaultmidpunct}
{\mcitedefaultendpunct}{\mcitedefaultseppunct}\relax
\EndOfBibitem
\bibitem[Bartlett and Musial(2007)Bartlett, and Musial]{Bartlett07}
Bartlett,~R.~J.; Musial,~M. Coupled-cluster theory in quantum chemistry.
  \emph{Reviews of Modern Physics} \textbf{2007}, \emph{79}, 291--352\relax
\mciteBstWouldAddEndPuncttrue
\mciteSetBstMidEndSepPunct{\mcitedefaultmidpunct}
{\mcitedefaultendpunct}{\mcitedefaultseppunct}\relax
\EndOfBibitem
\bibitem[Bartlett(2010)]{Bartlett10}
Bartlett,~R.~J. The coupled-cluster revolution. \emph{Molecular Physics}
  \textbf{2010}, \emph{108}, 2905--2920\relax
\mciteBstWouldAddEndPuncttrue
\mciteSetBstMidEndSepPunct{\mcitedefaultmidpunct}
{\mcitedefaultendpunct}{\mcitedefaultseppunct}\relax
\EndOfBibitem
\bibitem[Crawford and {Schaefer III}(2000)Crawford, and {Schaefer
  III}]{Crawford2000}
Crawford,~T.~D.; {Schaefer III},~H.~F. An introduction to coupled cluster
  theory for computational chemists. \emph{Reviews in Computational Chemistry}
  \textbf{2000}, \emph{14}, 33--136\relax
\mciteBstWouldAddEndPuncttrue
\mciteSetBstMidEndSepPunct{\mcitedefaultmidpunct}
{\mcitedefaultendpunct}{\mcitedefaultseppunct}\relax
\EndOfBibitem
\bibitem[Shavitt and Bartlett(2009)Shavitt, and Bartlett]{Shavitt2009}
Shavitt,~I.; Bartlett,~R.~J. \emph{Many-Body Methods in Chemistry and Physics:
  MBPT and Coupled-Cluster Theory}; Cambridge University Press: Cambridge,
  2009\relax
\mciteBstWouldAddEndPuncttrue
\mciteSetBstMidEndSepPunct{\mcitedefaultmidpunct}
{\mcitedefaultendpunct}{\mcitedefaultseppunct}\relax
\EndOfBibitem
\bibitem[Gauss(2000)]{Gauss00:properties}
Gauss,~J. In \emph{Modern Methods and Algorithms of Quantum Chemistry};
  Grotendorst,~J., Ed.; John von Neumann Institute for Computing: J{\"u}lich,
  2000; Vol.~1; pp 509--560\relax
\mciteBstWouldAddEndPuncttrue
\mciteSetBstMidEndSepPunct{\mcitedefaultmidpunct}
{\mcitedefaultendpunct}{\mcitedefaultseppunct}\relax
\EndOfBibitem
\bibitem[Hoffmann and {Schaefer III}(1986)Hoffmann, and {Schaefer
  III}]{Hoffmann86}
Hoffmann,~M.~R.; {Schaefer III},~H.~F. In \emph{Advances in Quantum Chemistry};
  L{\"o}wdin,~P.-O., Ed.; Academic Press: New York, 1986; Vol.~18; pp
  207--279\relax
\mciteBstWouldAddEndPuncttrue
\mciteSetBstMidEndSepPunct{\mcitedefaultmidpunct}
{\mcitedefaultendpunct}{\mcitedefaultseppunct}\relax
\EndOfBibitem
\bibitem[Noga and Bartlett(1987)Noga, and Bartlett]{Noga1987}
Noga,~J.; Bartlett,~R.~J. {The full CCSDT model for molecular electronic
  structure}. \emph{The Journal of Chemical Physics} \textbf{1987}, \emph{86},
  7041--7050\relax
\mciteBstWouldAddEndPuncttrue
\mciteSetBstMidEndSepPunct{\mcitedefaultmidpunct}
{\mcitedefaultendpunct}{\mcitedefaultseppunct}\relax
\EndOfBibitem
\bibitem[Lee \latin{et~al.}(1984)Lee, Kucharski, and Bartlett]{Lee1984}
Lee,~Y.~S.; Kucharski,~S.~A.; Bartlett,~R.~J. A coupled cluster approach with
  triple excitations. \emph{The Journal of Chemical Physics} \textbf{1984},
  \emph{81}, 5906--5912\relax
\mciteBstWouldAddEndPuncttrue
\mciteSetBstMidEndSepPunct{\mcitedefaultmidpunct}
{\mcitedefaultendpunct}{\mcitedefaultseppunct}\relax
\EndOfBibitem
\bibitem[Urban \latin{et~al.}(1985)Urban, Noga, Cole, and Bartlett]{Urban1985}
Urban,~M.; Noga,~J.; Cole,~S.~J.; Bartlett,~R.~J. Towards a full CCSDT model
  for electron correlation. \emph{The Journal of Chemical Physics}
  \textbf{1985}, \emph{83}, 4041--4046\relax
\mciteBstWouldAddEndPuncttrue
\mciteSetBstMidEndSepPunct{\mcitedefaultmidpunct}
{\mcitedefaultendpunct}{\mcitedefaultseppunct}\relax
\EndOfBibitem
\bibitem[Noga \latin{et~al.}(1987)Noga, Bartlett, and Urban]{Noga1987ccsdtn}
Noga,~J.; Bartlett,~R.~J.; Urban,~M. Towards a full CCSDT model for electron
  correlation. CCSDT-n models. \emph{Chemical Physics Letters} \textbf{1987},
  \emph{134}, 126--132\relax
\mciteBstWouldAddEndPuncttrue
\mciteSetBstMidEndSepPunct{\mcitedefaultmidpunct}
{\mcitedefaultendpunct}{\mcitedefaultseppunct}\relax
\EndOfBibitem
\bibitem[Raghavachari \latin{et~al.}(1989)Raghavachari, Trucks, Pople, and
  Head-Gordon]{Raghavachari1989}
Raghavachari,~K.; Trucks,~G.~W.; Pople,~J.~A.; Head-Gordon,~M. A fifth-order
  perturbation comparison of electron correlation theories. \emph{Chemical
  Physics Letters} \textbf{1989}, \emph{157}, 479--483\relax
\mciteBstWouldAddEndPuncttrue
\mciteSetBstMidEndSepPunct{\mcitedefaultmidpunct}
{\mcitedefaultendpunct}{\mcitedefaultseppunct}\relax
\EndOfBibitem
\bibitem[Stanton(1997)]{Stanton1997}
Stanton,~J.~F. Why CCSD (T) works: a different perspective. \emph{Chemical
  Physics Letters} \textbf{1997}, \emph{281}, 130--134\relax
\mciteBstWouldAddEndPuncttrue
\mciteSetBstMidEndSepPunct{\mcitedefaultmidpunct}
{\mcitedefaultendpunct}{\mcitedefaultseppunct}\relax
\EndOfBibitem
\bibitem[Crawford and Stanton(1998)Crawford, and Stanton]{Crawford98:aT}
Crawford,~T.~D.; Stanton,~J.~F. Investigation of an asymmetric
  triple-excitation correction for coupled cluster energies.
  \emph{International Journal of Quantum Chemistry Symp.} \textbf{1998},
  \emph{70}, 601--611\relax
\mciteBstWouldAddEndPuncttrue
\mciteSetBstMidEndSepPunct{\mcitedefaultmidpunct}
{\mcitedefaultendpunct}{\mcitedefaultseppunct}\relax
\EndOfBibitem
\bibitem[Kucharski and Bartlett(1998)Kucharski, and Bartlett]{Kucharski98}
Kucharski,~S.~A.; Bartlett,~R.~J. Noniterative energy corrections through
  fifth-order to the coupled cluster singles and doubles method. \emph{The
  Journal of Chemical Physics} \textbf{1998}, \emph{108}, 5243\relax
\mciteBstWouldAddEndPuncttrue
\mciteSetBstMidEndSepPunct{\mcitedefaultmidpunct}
{\mcitedefaultendpunct}{\mcitedefaultseppunct}\relax
\EndOfBibitem
\bibitem[Kowalski and Piecuch(2000)Kowalski, and Piecuch]{Kowalski2000}
Kowalski,~K.; Piecuch,~P. The method of moments of coupled-cluster equations
  and the renormalized CCSD [T], CCSD (T), CCSD (TQ), and CCSDT (Q) approaches.
  \emph{The Journal of Chemical Physics} \textbf{2000}, \emph{113},
  18--35\relax
\mciteBstWouldAddEndPuncttrue
\mciteSetBstMidEndSepPunct{\mcitedefaultmidpunct}
{\mcitedefaultendpunct}{\mcitedefaultseppunct}\relax
\EndOfBibitem
\bibitem[Piecuch \latin{et~al.}(2002)Piecuch, Kucharski, Kowalski, and
  Musia{\l}]{Piecuch2002}
Piecuch,~P.; Kucharski,~S.~A.; Kowalski,~K.; Musia{\l},~M. Efficient computer
  implementation of the renormalized coupled-cluster methods: the r-ccsd [t],
  r-ccsd (t), cr-ccsd [t], and cr-ccsd (t) approaches. \emph{Computer Physics
  Communications} \textbf{2002}, \emph{149}, 71--96\relax
\mciteBstWouldAddEndPuncttrue
\mciteSetBstMidEndSepPunct{\mcitedefaultmidpunct}
{\mcitedefaultendpunct}{\mcitedefaultseppunct}\relax
\EndOfBibitem
\bibitem[Christiansen \latin{et~al.}(1995)Christiansen, Koch, and
  J{\o}rgensen]{Christiansen1995CC3}
Christiansen,~O.; Koch,~H.; J{\o}rgensen,~P. Response functions in the CC3
  iterative triple excitation model. \emph{The Journal of Chemical Physics}
  \textbf{1995}, \emph{103}, 7429--7441\relax
\mciteBstWouldAddEndPuncttrue
\mciteSetBstMidEndSepPunct{\mcitedefaultmidpunct}
{\mcitedefaultendpunct}{\mcitedefaultseppunct}\relax
\EndOfBibitem
\bibitem[Koch \latin{et~al.}(1997)Koch, Christiansen, Sanchez~de Mer{\'a}s,
  Helgaker, \latin{et~al.} others]{Koch1997}
Koch,~H.; Christiansen,~O.; Sanchez~de Mer{\'a}s,~A.~M.; Helgaker,~T.; others
  {The CC3 model: An iterative coupled cluster approach including connected
  triples}. \emph{The Journal of Chemical Physics} \textbf{1997}, \emph{106},
  1808--1818\relax
\mciteBstWouldAddEndPuncttrue
\mciteSetBstMidEndSepPunct{\mcitedefaultmidpunct}
{\mcitedefaultendpunct}{\mcitedefaultseppunct}\relax
\EndOfBibitem
\bibitem[Christiansen \latin{et~al.}(1998)Christiansen, Gauss, and
  Stanton]{Christiansen1998triple}
Christiansen,~O.; Gauss,~J.; Stanton,~J.~F. The effect of triple excitations in
  coupled cluster calculations of frequency-dependent polarizabilities.
  \emph{Chemical Physics Letters} \textbf{1998}, \emph{292}, 437--446\relax
\mciteBstWouldAddEndPuncttrue
\mciteSetBstMidEndSepPunct{\mcitedefaultmidpunct}
{\mcitedefaultendpunct}{\mcitedefaultseppunct}\relax
\EndOfBibitem
\bibitem[Gauss \latin{et~al.}(1998)Gauss, Christiansen, and Stanton]{Gauss1998}
Gauss,~J.; Christiansen,~O.; Stanton,~J.~F. Triple excitation effects in
  coupled-cluster calculations of frequency-dependent hyperpolarizabilities.
  \emph{Chemical Physics Letters} \textbf{1998}, \emph{296}, 117--124\relax
\mciteBstWouldAddEndPuncttrue
\mciteSetBstMidEndSepPunct{\mcitedefaultmidpunct}
{\mcitedefaultendpunct}{\mcitedefaultseppunct}\relax
\EndOfBibitem
\bibitem[Olsen and J{\o}rgensen(1985)Olsen, and J{\o}rgensen]{Olsen1985}
Olsen,~J.; J{\o}rgensen,~P. {Linear and nonlinear response functions for an
  exact state and for an MCSCF state}. \emph{The Journal of Chemical Physics}
  \textbf{1985}, \emph{82}, 3235--3264\relax
\mciteBstWouldAddEndPuncttrue
\mciteSetBstMidEndSepPunct{\mcitedefaultmidpunct}
{\mcitedefaultendpunct}{\mcitedefaultseppunct}\relax
\EndOfBibitem
\bibitem[Sekino and Bartlett(1984)Sekino, and Bartlett]{Sekino1984}
Sekino,~H.; Bartlett,~R.~J. A linear response, coupled-cluster theory for
  excitation energy. \emph{International Journal of Quantum Chemistry}
  \textbf{1984}, \emph{26}, 255--265\relax
\mciteBstWouldAddEndPuncttrue
\mciteSetBstMidEndSepPunct{\mcitedefaultmidpunct}
{\mcitedefaultendpunct}{\mcitedefaultseppunct}\relax
\EndOfBibitem
\bibitem[Helgaker \latin{et~al.}(2012)Helgaker, Coriani, J{\o}rgensen,
  Kristensen, Olsen, and Ruud]{Helgaker12}
Helgaker,~T.; Coriani,~S.; J{\o}rgensen,~P.; Kristensen,~K.; Olsen,~J.;
  Ruud,~K. Recent Advances in Wave Function-Based Methods of Molecular-Property
  Calculations. \emph{Chemical Reviews} \textbf{2012}, \emph{112},
  543--631\relax
\mciteBstWouldAddEndPuncttrue
\mciteSetBstMidEndSepPunct{\mcitedefaultmidpunct}
{\mcitedefaultendpunct}{\mcitedefaultseppunct}\relax
\EndOfBibitem
\bibitem[Norman \latin{et~al.}(2018)Norman, Ruud, and Saue]{Norman18}
Norman,~P.; Ruud,~K.; Saue,~T. \emph{Principles and Practices of Molecular
  Properties: Theory, Modeling, and Simulations}; John Wiley and Sons: 111
  River Street, Hoboken, NJ 07030, USA, 2018\relax
\mciteBstWouldAddEndPuncttrue
\mciteSetBstMidEndSepPunct{\mcitedefaultmidpunct}
{\mcitedefaultendpunct}{\mcitedefaultseppunct}\relax
\EndOfBibitem
\bibitem[Goings \latin{et~al.}(2018)Goings, Lestrange, and Li]{Goings2018}
Goings,~J.~J.; Lestrange,~P.~J.; Li,~X. Real-time time-dependent electronic
  structure theory. \emph{Wiley Interdisciplinary Reviews: Computational
  Molecular Science} \textbf{2018}, \emph{8}, e1341\relax
\mciteBstWouldAddEndPuncttrue
\mciteSetBstMidEndSepPunct{\mcitedefaultmidpunct}
{\mcitedefaultendpunct}{\mcitedefaultseppunct}\relax
\EndOfBibitem
\bibitem[Li \latin{et~al.}(2020)Li, Govind, Isborn, DePrince~III, and
  Lopata]{Li2020}
Li,~X.; Govind,~N.; Isborn,~C.; DePrince~III,~A.~E.; Lopata,~K. Real-time
  time-dependent electronic structure theory. \emph{Chemical Reviews}
  \textbf{2020}, \emph{120}, 9951--9993\relax
\mciteBstWouldAddEndPuncttrue
\mciteSetBstMidEndSepPunct{\mcitedefaultmidpunct}
{\mcitedefaultendpunct}{\mcitedefaultseppunct}\relax
\EndOfBibitem
\bibitem[Ofstad \latin{et~al.}(2023)Ofstad, Aurbakken, Sch{\o}yen, Kristiansen,
  Kvaal, and Pedersen]{Ofstad2023_review}
Ofstad,~B.~S.; Aurbakken,~E.; Sch{\o}yen,~{\O}.~S.; Kristiansen,~H.~E.;
  Kvaal,~S.; Pedersen,~T.~B. Time-dependent coupled-cluster theory. \emph{Wiley
  Interdisciplinary Reviews: Computational Molecular Science} \textbf{2023},
  \emph{13}, e1666\relax
\mciteBstWouldAddEndPuncttrue
\mciteSetBstMidEndSepPunct{\mcitedefaultmidpunct}
{\mcitedefaultendpunct}{\mcitedefaultseppunct}\relax
\EndOfBibitem
\bibitem[Huber and Klamroth(2011)Huber, and Klamroth]{Huber2011}
Huber,~C.; Klamroth,~T. Explicitly time-dependent coupled cluster singles
  doubles calculations of laser-driven many-electron dynamics. \emph{The
  Journal of Chemical Physics} \textbf{2011}, \emph{134}, 054113\relax
\mciteBstWouldAddEndPuncttrue
\mciteSetBstMidEndSepPunct{\mcitedefaultmidpunct}
{\mcitedefaultendpunct}{\mcitedefaultseppunct}\relax
\EndOfBibitem
\bibitem[Kvaal(2012)]{Kvaal2012}
Kvaal,~S. Ab initio quantum dynamics using coupled-cluster. \emph{The Journal
  of Chemical Physics} \textbf{2012}, \emph{136}, 194109\relax
\mciteBstWouldAddEndPuncttrue
\mciteSetBstMidEndSepPunct{\mcitedefaultmidpunct}
{\mcitedefaultendpunct}{\mcitedefaultseppunct}\relax
\EndOfBibitem
\bibitem[Sato \latin{et~al.}(2018)Sato, Pathak, Orimo, and Ishikawa]{Sato2018}
Sato,~T.; Pathak,~H.; Orimo,~Y.; Ishikawa,~K.~L. Time-dependent optimized
  coupled-cluster method for multielectron dynamics. \emph{The Journal of
  Chemical Physics} \textbf{2018}, \emph{148}, 051101\relax
\mciteBstWouldAddEndPuncttrue
\mciteSetBstMidEndSepPunct{\mcitedefaultmidpunct}
{\mcitedefaultendpunct}{\mcitedefaultseppunct}\relax
\EndOfBibitem
\bibitem[Pedersen and Kvaal(2019)Pedersen, and Kvaal]{Pedersen2019}
Pedersen,~T.~B.; Kvaal,~S. Symplectic integration and physical interpretation
  of time-dependent coupled-cluster theory. \emph{The Journal of Chemical
  Physics} \textbf{2019}, \emph{150}, 144106\relax
\mciteBstWouldAddEndPuncttrue
\mciteSetBstMidEndSepPunct{\mcitedefaultmidpunct}
{\mcitedefaultendpunct}{\mcitedefaultseppunct}\relax
\EndOfBibitem
\bibitem[Nascimento and DePrince~III(2016)Nascimento, and
  DePrince~III]{Nascimento2016}
Nascimento,~D.~R.; DePrince~III,~A.~E. Linear absorption spectra from
  explicitly time-dependent equation-of-motion coupled-cluster theory.
  \emph{Journal of Chemical Theory and Computation} \textbf{2016}, \emph{12},
  5834--5840\relax
\mciteBstWouldAddEndPuncttrue
\mciteSetBstMidEndSepPunct{\mcitedefaultmidpunct}
{\mcitedefaultendpunct}{\mcitedefaultseppunct}\relax
\EndOfBibitem
\bibitem[Nascimento and DePrince~III(2019)Nascimento, and
  DePrince~III]{Nascimento2019}
Nascimento,~D.~R.; DePrince~III,~A.~E. A general time-domain formulation of
  equation-of-motion coupled-cluster theory for linear spectroscopy. \emph{The
  Journal of Chemical Physics} \textbf{2019}, \emph{151}, 204107\relax
\mciteBstWouldAddEndPuncttrue
\mciteSetBstMidEndSepPunct{\mcitedefaultmidpunct}
{\mcitedefaultendpunct}{\mcitedefaultseppunct}\relax
\EndOfBibitem
\bibitem[Pedersen \latin{et~al.}(2021)Pedersen, Kristiansen, Bodenstein, Kvaal,
  and Sch{\o}yen]{Pedersen2021}
Pedersen,~T.~B.; Kristiansen,~H.~E.; Bodenstein,~T.; Kvaal,~S.;
  Sch{\o}yen,~{\O}.~S. Interpretation of Coupled-Cluster Many-Electron Dynamics
  in Terms of Stationary States. \emph{Journal of Chemical Theory and
  Computation} \textbf{2021}, \emph{17}, 388--404\relax
\mciteBstWouldAddEndPuncttrue
\mciteSetBstMidEndSepPunct{\mcitedefaultmidpunct}
{\mcitedefaultendpunct}{\mcitedefaultseppunct}\relax
\EndOfBibitem
\bibitem[Wang \latin{et~al.}(2022)Wang, Peyton, and Crawford]{Wang2022}
Wang,~Z.; Peyton,~B.~G.; Crawford,~T.~D. Accelerating real-time coupled cluster
  methods with single-precision arithmetic and adaptive numerical integration.
  \emph{Journal of Chemical Theory and Computation} \textbf{2022}, \emph{18},
  5479--5491\relax
\mciteBstWouldAddEndPuncttrue
\mciteSetBstMidEndSepPunct{\mcitedefaultmidpunct}
{\mcitedefaultendpunct}{\mcitedefaultseppunct}\relax
\EndOfBibitem
\bibitem[Peyton \latin{et~al.}(2023)Peyton, Wang, and Crawford]{Peyton2023}
Peyton,~B.~G.; Wang,~Z.; Crawford,~T.~D. Reduced Scaling Real-Time Coupled
  Cluster Theory,. \emph{Journal of Physical Chemistry A} \textbf{2023},
  \emph{127}, 8486--8499\relax
\mciteBstWouldAddEndPuncttrue
\mciteSetBstMidEndSepPunct{\mcitedefaultmidpunct}
{\mcitedefaultendpunct}{\mcitedefaultseppunct}\relax
\EndOfBibitem
\bibitem[Hald and J{\o}rgensen(2002)Hald, and J{\o}rgensen]{Hald2002}
Hald,~K.; J{\o}rgensen,~P. Calculation of first-order one-electron properties
  using the coupled-cluster approximate triples model CC3. \emph{Physical
  Chemistry Chemical Physics} \textbf{2002}, \emph{4}, 5221--5226\relax
\mciteBstWouldAddEndPuncttrue
\mciteSetBstMidEndSepPunct{\mcitedefaultmidpunct}
{\mcitedefaultendpunct}{\mcitedefaultseppunct}\relax
\EndOfBibitem
\bibitem[Hald \latin{et~al.}(2003)Hald, Paw{\l}owski, J{\o}rgensen, and
  H{\"a}ttig]{Hald2003}
Hald,~K.; Paw{\l}owski,~F.; J{\o}rgensen,~P.; H{\"a}ttig,~C. Calculation of
  frequency-dependent polarizabilities using the approximate coupled-cluster
  triples model CC3. \emph{The Journal of Chemical Physics} \textbf{2003},
  \emph{118}, 1292--1300\relax
\mciteBstWouldAddEndPuncttrue
\mciteSetBstMidEndSepPunct{\mcitedefaultmidpunct}
{\mcitedefaultendpunct}{\mcitedefaultseppunct}\relax
\EndOfBibitem
\bibitem[Pathak \latin{et~al.}(2021)Pathak, Sato, and Ishikawa]{Pathak2021}
Pathak,~H.; Sato,~T.; Ishikawa,~K.~L. {Time-dependent optimized coupled-cluster
  method for multielectron dynamics. IV. Approximate consideration of the
  triple excitation amplitudes}. \emph{The Journal of Chemical Physics}
  \textbf{2021}, \emph{154}, 234104\relax
\mciteBstWouldAddEndPuncttrue
\mciteSetBstMidEndSepPunct{\mcitedefaultmidpunct}
{\mcitedefaultendpunct}{\mcitedefaultseppunct}\relax
\EndOfBibitem
\bibitem[Pathak \latin{et~al.}(2022)Pathak, Sato, and Ishikawa]{Pathak2022}
Pathak,~H.; Sato,~T.; Ishikawa,~K.~L. Time-dependent optimized coupled-cluster
  method with doubles and perturbative triples for first principles simulation
  of multielectron dynamics. \emph{Frontiers in Chemistry} \textbf{2022},
  \emph{10}, 982120\relax
\mciteBstWouldAddEndPuncttrue
\mciteSetBstMidEndSepPunct{\mcitedefaultmidpunct}
{\mcitedefaultendpunct}{\mcitedefaultseppunct}\relax
\EndOfBibitem
\bibitem[Perrone and Kao(1975)Perrone, and Kao]{Perrone1975}
Perrone,~N.; Kao,~R. A general finite difference method for arbitrary meshes.
  \emph{Computers \& Structures} \textbf{1975}, \emph{5}, 45--57\relax
\mciteBstWouldAddEndPuncttrue
\mciteSetBstMidEndSepPunct{\mcitedefaultmidpunct}
{\mcitedefaultendpunct}{\mcitedefaultseppunct}\relax
\EndOfBibitem
\bibitem[Ding \latin{et~al.}(2013)Ding, Van~Kuiken, Eichinger, and
  Li]{Ding2013}
Ding,~F.; Van~Kuiken,~B.~E.; Eichinger,~B.~E.; Li,~X. An efficient method for
  calculating dynamical hyperpolarizabilities using real-time time-dependent
  density functional theory. \emph{The Journal of Chemical Physics}
  \textbf{2013}, \emph{138}, 064104\relax
\mciteBstWouldAddEndPuncttrue
\mciteSetBstMidEndSepPunct{\mcitedefaultmidpunct}
{\mcitedefaultendpunct}{\mcitedefaultseppunct}\relax
\EndOfBibitem
\bibitem[Ofstad \latin{et~al.}(2023)Ofstad, Kristiansen, Aurbakken, Sch{\o}yen,
  Kvaal, and Pedersen]{Ofstad2023}
Ofstad,~B.~S.; Kristiansen,~H.~E.; Aurbakken,~E.; Sch{\o}yen,~{\O}.~S.;
  Kvaal,~S.; Pedersen,~T.~B. Adiabatic extraction of nonlinear optical
  properties from real-time time-dependent electronic-structure theory.
  \emph{The Journal of Chemical Physics} \textbf{2023}, \emph{158},
  154102\relax
\mciteBstWouldAddEndPuncttrue
\mciteSetBstMidEndSepPunct{\mcitedefaultmidpunct}
{\mcitedefaultendpunct}{\mcitedefaultseppunct}\relax
\EndOfBibitem
\bibitem[Bruner \latin{et~al.}(2016)Bruner, LaMaster, and Lopata]{Bruner2016}
Bruner,~A.; LaMaster,~D.; Lopata,~K. Accelerated broadband spectra using
  transition dipole decomposition and Pad{\'e} approximants. \emph{Journal of
  Chemical Theory and Computation} \textbf{2016}, \emph{12}, 3741--3750\relax
\mciteBstWouldAddEndPuncttrue
\mciteSetBstMidEndSepPunct{\mcitedefaultmidpunct}
{\mcitedefaultendpunct}{\mcitedefaultseppunct}\relax
\EndOfBibitem
\bibitem[Harris \latin{et~al.}(2020)Harris, Millman, Van Der~Walt, Gommers,
  Virtanen, Cournapeau, Wieser, Taylor, Berg, Smith, \latin{et~al.}
  others]{Harris2020}
Harris,~C.~R.; Millman,~K.~J.; Van Der~Walt,~S.~J.; Gommers,~R.; Virtanen,~P.;
  Cournapeau,~D.; Wieser,~E.; Taylor,~J.; Berg,~S.; Smith,~N.~J. \latin{et~al.}
   Array programming with NumPy. \emph{Nature} \textbf{2020}, \emph{585},
  357--362\relax
\mciteBstWouldAddEndPuncttrue
\mciteSetBstMidEndSepPunct{\mcitedefaultmidpunct}
{\mcitedefaultendpunct}{\mcitedefaultseppunct}\relax
\EndOfBibitem
\bibitem[Paszke \latin{et~al.}(2019)Paszke, Gross, Massa, Lerer, Bradbury,
  Chanan, Killeen, Lin, Gimelshein, Antiga, \latin{et~al.} others]{Paszke2019}
Paszke,~A.; Gross,~S.; Massa,~F.; Lerer,~A.; Bradbury,~J.; Chanan,~G.;
  Killeen,~T.; Lin,~Z.; Gimelshein,~N.; Antiga,~L. \latin{et~al.}  Pytorch: An
  imperative style, high-performance deep learning library. \emph{Advances in
  Neural Information Processing Systems} \textbf{2019}, \emph{32},
  8024--8035\relax
\mciteBstWouldAddEndPuncttrue
\mciteSetBstMidEndSepPunct{\mcitedefaultmidpunct}
{\mcitedefaultendpunct}{\mcitedefaultseppunct}\relax
\EndOfBibitem
\bibitem[Smith and Gray(2018)Smith, and Gray]{Smith2018}
Smith,~D. G.~A.; Gray,~J. {opt\textunderscore einsum -- A Python package for
  optimizing contraction order for einsum-like expressions}. \emph{Journal of
  Open Source Software} \textbf{2018}, \emph{3}, 753\relax
\mciteBstWouldAddEndPuncttrue
\mciteSetBstMidEndSepPunct{\mcitedefaultmidpunct}
{\mcitedefaultendpunct}{\mcitedefaultseppunct}\relax
\EndOfBibitem
\bibitem[Dunning~Jr(1989)]{Dunning1989}
Dunning~Jr,~T.~H. Gaussian basis sets for use in correlated molecular
  calculations. I. The atoms boron through neon and hydrogen. \emph{The Journal
  of Chemical Physics} \textbf{1989}, \emph{90}, 1007--1023\relax
\mciteBstWouldAddEndPuncttrue
\mciteSetBstMidEndSepPunct{\mcitedefaultmidpunct}
{\mcitedefaultendpunct}{\mcitedefaultseppunct}\relax
\EndOfBibitem
\bibitem[Virtanen \latin{et~al.}(2020)Virtanen, Gommers, Oliphant, Haberland,
  Reddy, Cournapeau, Burovski, Peterson, Weckesser, Bright, \latin{et~al.}
  others]{Virtanen2020}
Virtanen,~P.; Gommers,~R.; Oliphant,~T.~E.; Haberland,~M.; Reddy,~T.;
  Cournapeau,~D.; Burovski,~E.; Peterson,~P.; Weckesser,~W.; Bright,~J.
  \latin{et~al.}  SciPy 1.0: fundamental algorithms for scientific computing in
  Python. \emph{Nature Methods} \textbf{2020}, \emph{17}, 261--272\relax
\mciteBstWouldAddEndPuncttrue
\mciteSetBstMidEndSepPunct{\mcitedefaultmidpunct}
{\mcitedefaultendpunct}{\mcitedefaultseppunct}\relax
\EndOfBibitem
\bibitem[Smith \latin{et~al.}(2020)Smith, Burns, Simmonett, Parrish, Schieber,
  Galvelis, Kraus, Kruse, Di~Remigio, Alenaizan, \latin{et~al.}
  others]{Smith2020}
Smith,~D.~G.; Burns,~L.~A.; Simmonett,~A.~C.; Parrish,~R.~M.; Schieber,~M.~C.;
  Galvelis,~R.; Kraus,~P.; Kruse,~H.; Di~Remigio,~R.; Alenaizan,~A.
  \latin{et~al.}  PSI4 1.4: Open-source software for high-throughput quantum
  chemistry. \emph{The Journal of Chemical Physics} \textbf{2020}, \emph{152},
  184108\relax
\mciteBstWouldAddEndPuncttrue
\mciteSetBstMidEndSepPunct{\mcitedefaultmidpunct}
{\mcitedefaultendpunct}{\mcitedefaultseppunct}\relax
\EndOfBibitem
\bibitem[Matthews \latin{et~al.}(2020)Matthews, Cheng, Harding, Lipparini,
  Stopkowicz, Jagau, Szalay, Gauss, and Stanton]{Matthews2020}
Matthews,~D.~A.; Cheng,~L.; Harding,~M.~E.; Lipparini,~F.; Stopkowicz,~S.;
  Jagau,~T.-C.; Szalay,~P.~G.; Gauss,~J.; Stanton,~J.~F. Coupled-cluster
  techniques for computational chemistry: The CFOUR program package. \emph{The
  Journal of Chemical Physics} \textbf{2020}, \emph{152}, 214108\relax
\mciteBstWouldAddEndPuncttrue
\mciteSetBstMidEndSepPunct{\mcitedefaultmidpunct}
{\mcitedefaultendpunct}{\mcitedefaultseppunct}\relax
\EndOfBibitem
\bibitem[Pedersen \latin{et~al.}(2001)Pedersen, Fern{\'a}ndez, and
  Koch]{Pedersen2001}
Pedersen,~T.~B.; Fern{\'a}ndez,~B.; Koch,~H. Gauge invariant coupled cluster
  response theory using optimized nonorthogonal orbitals. \emph{The Journal of
  Chemical Physics} \textbf{2001}, \emph{114}, 6983--6993\relax
\mciteBstWouldAddEndPuncttrue
\mciteSetBstMidEndSepPunct{\mcitedefaultmidpunct}
{\mcitedefaultendpunct}{\mcitedefaultseppunct}\relax
\EndOfBibitem
\bibitem[Woon and Dunning~Jr(1993)Woon, and Dunning~Jr]{Woon1993}
Woon,~D.~E.; Dunning~Jr,~T.~H. Gaussian basis sets for use in correlated
  molecular calculations. III. The atoms aluminum through argon. \emph{The
  Journal of Chemical Physics} \textbf{1993}, \emph{98}, 1358--1371\relax
\mciteBstWouldAddEndPuncttrue
\mciteSetBstMidEndSepPunct{\mcitedefaultmidpunct}
{\mcitedefaultendpunct}{\mcitedefaultseppunct}\relax
\EndOfBibitem
\bibitem[Woon and Dunning~Jr(1994)Woon, and Dunning~Jr]{Woon1994}
Woon,~D.~E.; Dunning~Jr,~T.~H. Gaussian basis sets for use in correlated
  molecular calculations. IV. Calculation of static electrical response
  properties. \emph{The Journal of Chemical Physics} \textbf{1994}, \emph{100},
  2975--2988\relax
\mciteBstWouldAddEndPuncttrue
\mciteSetBstMidEndSepPunct{\mcitedefaultmidpunct}
{\mcitedefaultendpunct}{\mcitedefaultseppunct}\relax
\EndOfBibitem
\bibitem[Crawford \latin{et~al.}()Crawford, Peyton, Wang, and Madriaga]{pycc}
Crawford,~T.~D.; Peyton,~B.~G.; Wang,~Z.; Madriaga,~J.~M.
  http://github.com/CrawfordGroup/pycc\relax
\mciteBstWouldAddEndPuncttrue
\mciteSetBstMidEndSepPunct{\mcitedefaultmidpunct}
{\mcitedefaultendpunct}{\mcitedefaultseppunct}\relax
\EndOfBibitem
\bibitem[Butcher(1996)]{Butcher1996}
Butcher,~J.~C. A history of Runge-Kutta methods. \emph{Applied Numerical
  Mathematics} \textbf{1996}, \emph{20}, 247--260\relax
\mciteBstWouldAddEndPuncttrue
\mciteSetBstMidEndSepPunct{\mcitedefaultmidpunct}
{\mcitedefaultendpunct}{\mcitedefaultseppunct}\relax
\EndOfBibitem
\bibitem[Pokhilko \latin{et~al.}(2018)Pokhilko, Epifanovsky, and
  Krylov]{Pokhilko2018}
Pokhilko,~P.; Epifanovsky,~E.; Krylov,~A.~I. Double precision is not needed for
  many-body calculations: Emergent conventional wisdom. \emph{Journal of
  Chemical Theory and Computation} \textbf{2018}, \emph{14}, 4088--4096\relax
\mciteBstWouldAddEndPuncttrue
\mciteSetBstMidEndSepPunct{\mcitedefaultmidpunct}
{\mcitedefaultendpunct}{\mcitedefaultseppunct}\relax
\EndOfBibitem
\bibitem[Kristiansen \latin{et~al.}(2022)Kristiansen, Ofstad, Hauge, Aurbakken,
  Sch{\o}yen, Kvaal, and Pedersen]{Kristiansen2022}
Kristiansen,~H.~E.; Ofstad,~B.~S.; Hauge,~E.; Aurbakken,~E.;
  Sch{\o}yen,~{\O}.~S.; Kvaal,~S.; Pedersen,~T.~B. {Linear and nonlinear
  optical properties from TDOMP2 theory}. \emph{Journal of Chemical Theory and
  Computation} \textbf{2022}, \emph{18}, 3687--3702\relax
\mciteBstWouldAddEndPuncttrue
\mciteSetBstMidEndSepPunct{\mcitedefaultmidpunct}
{\mcitedefaultendpunct}{\mcitedefaultseppunct}\relax
\EndOfBibitem
\bibitem[Paul \latin{et~al.}(2020)Paul, Myhre, and Koch]{Paul2020}
Paul,~A.~C.; Myhre,~R.~H.; Koch,~H. New and efficient implementation of CC3.
  \emph{Journal of Chemical Theory and Computation} \textbf{2020}, \emph{17},
  117--126\relax
\mciteBstWouldAddEndPuncttrue
\mciteSetBstMidEndSepPunct{\mcitedefaultmidpunct}
{\mcitedefaultendpunct}{\mcitedefaultseppunct}\relax
\EndOfBibitem
\bibitem[Pathak \latin{et~al.}(2020)Pathak, Sato, and Ishikawa]{Pathak2020}
Pathak,~H.; Sato,~T.; Ishikawa,~K.~L. Time-dependent optimized coupled-cluster
  method for multielectron dynamics. {III}. {A} second-order many-body
  perturbation approximation. \emph{Journal of Chemical Physics} \textbf{2020},
  \emph{153}, 034110\relax
\mciteBstWouldAddEndPuncttrue
\mciteSetBstMidEndSepPunct{\mcitedefaultmidpunct}
{\mcitedefaultendpunct}{\mcitedefaultseppunct}\relax
\EndOfBibitem
\bibitem[Pedersen \latin{et~al.}(1999)Pedersen, Koch, and
  H{\"a}ttig]{Pedersen1999}
Pedersen,~T.~B.; Koch,~H.; H{\"a}ttig,~C. Gauge invariant coupled cluster
  response theory. \emph{The Journal of Chemical Physics} \textbf{1999},
  \emph{110}, 8318--8327\relax
\mciteBstWouldAddEndPuncttrue
\mciteSetBstMidEndSepPunct{\mcitedefaultmidpunct}
{\mcitedefaultendpunct}{\mcitedefaultseppunct}\relax
\EndOfBibitem
\bibitem[Ofstad \latin{et~al.}(2023)Ofstad, Wibowo-Teale, Kristiansen,
  Aurbakken, Kitsaras, Sch{\o}yen, Hauge, Irons, Kvaal, Stopkowicz,
  Wibowo-Teale, and Pedersen]{Ofstad2023_magnetic}
Ofstad,~B.~S.; Wibowo-Teale,~M.; Kristiansen,~H.~E.; Aurbakken,~E.;
  Kitsaras,~M.~P.; Sch{\o}yen,~{\O}.~S.; Hauge,~E.; Irons,~T. J.~P.; Kvaal,~S.;
  Stopkowicz,~S. \latin{et~al.}  Magnetic optical rotation from real-time
  simulations in finite magnetic fields. \emph{The Journal of Chemical Physics}
  \textbf{2023}, \emph{159}, 204109\relax
\mciteBstWouldAddEndPuncttrue
\mciteSetBstMidEndSepPunct{\mcitedefaultmidpunct}
{\mcitedefaultendpunct}{\mcitedefaultseppunct}\relax
\EndOfBibitem
\bibitem[K{\"o}hn and Olsen(2005)K{\"o}hn, and Olsen]{Kohn2005}
K{\"o}hn,~A.; Olsen,~J. Orbital-optimized coupled-cluster theory does not
  reproduce the full configuration-interaction limit. \emph{The Journal of
  Chemical Physics} \textbf{2005}, \emph{122}, 084116\relax
\mciteBstWouldAddEndPuncttrue
\mciteSetBstMidEndSepPunct{\mcitedefaultmidpunct}
{\mcitedefaultendpunct}{\mcitedefaultseppunct}\relax
\EndOfBibitem
\bibitem[Myhre(2018)]{Myhre2018}
Myhre,~R.~H. Demonstrating that the nonorthogonal orbital optimized coupled
  cluster model converges to full configuration interaction. \emph{The Journal
  of Chemical Physics} \textbf{2018}, \emph{148}, 094110\relax
\mciteBstWouldAddEndPuncttrue
\mciteSetBstMidEndSepPunct{\mcitedefaultmidpunct}
{\mcitedefaultendpunct}{\mcitedefaultseppunct}\relax
\EndOfBibitem
\bibitem[{Aurbakken, E. and Fredly, K. H. and Kristiansen, H. E. and Kvaal, S.
  and Myhre, R. H. and Ofstad, B. S. and Pedersen, T. B. and Sch{\o}yen, {\O}.
  S. and Sutterud, H. and Winther-Larsen, S. G.}(2024)]{HyQD}
{Aurbakken, E. and Fredly, K. H. and Kristiansen, H. E. and Kvaal, S. and
  Myhre, R. H. and Ofstad, B. S. and Pedersen, T. B. and Sch{\o}yen, {\O}. S.
  and Sutterud, H. and Winther-Larsen, S. G.} {HyQD: Hylleraas Quantum
  Dynamics}. 2024; \url{https://github.com/HyQD}, (accessed May 24, 2024)\relax
\mciteBstWouldAddEndPuncttrue
\mciteSetBstMidEndSepPunct{\mcitedefaultmidpunct}
{\mcitedefaultendpunct}{\mcitedefaultseppunct}\relax
\EndOfBibitem
\bibitem[Larsen \latin{et~al.}(1999)Larsen, Olsen, H{\"a}ttig, J{\o}rgensen,
  Christiansen, and Gauss]{Larsen1999}
Larsen,~H.; Olsen,~J.; H{\"a}ttig,~C.; J{\o}rgensen,~P.; Christiansen,~O.;
  Gauss,~J. Polarizabilities and first hyperpolarizabilities of HF, Ne, and BH
  from full configuration interaction and coupled cluster calculations.
  \emph{The Journal of Chemical Physics} \textbf{1999}, \emph{111},
  1917--1925\relax
\mciteBstWouldAddEndPuncttrue
\mciteSetBstMidEndSepPunct{\mcitedefaultmidpunct}
{\mcitedefaultendpunct}{\mcitedefaultseppunct}\relax
\EndOfBibitem
\bibitem[Hirschfelder \latin{et~al.}(1964)Hirschfelder, Brown, and
  Epstein]{Hirschfelder1964}
Hirschfelder,~J.~O.; Brown,~W.~B.; Epstein,~S.~T. In \emph{Advances in Quantum
  Chemistry}; L{\"o}wdin,~P.-O., Ed.; Academic Press: New York, 1964; Vol.~1;
  pp 255--374\relax
\mciteBstWouldAddEndPuncttrue
\mciteSetBstMidEndSepPunct{\mcitedefaultmidpunct}
{\mcitedefaultendpunct}{\mcitedefaultseppunct}\relax
\EndOfBibitem
\bibitem[H{\"a}ttig \latin{et~al.}(1997)H{\"a}ttig, Christiansen, and
  J{\o}rgensen]{Hattig1997}
H{\"a}ttig,~C.; Christiansen,~O.; J{\o}rgensen,~P. Cauchy moments and
  dispersion coefficients using coupled cluster linear response theory.
  \emph{The Journal of Chemical Physics} \textbf{1997}, \emph{107},
  10592--10598\relax
\mciteBstWouldAddEndPuncttrue
\mciteSetBstMidEndSepPunct{\mcitedefaultmidpunct}
{\mcitedefaultendpunct}{\mcitedefaultseppunct}\relax
\EndOfBibitem
\bibitem[Kristiansen \latin{et~al.}(2020)Kristiansen, Sch{\o}yen, Kvaal, and
  Pedersen]{Kristiansen2020}
Kristiansen,~H.~E.; Sch{\o}yen,~{\O}.~S.; Kvaal,~S.; Pedersen,~T.~B. Numerical
  stability of time-dependent coupled-cluster methods for many-electron
  dynamics in intense laser pulses. \emph{The Journal of Chemical Physics}
  \textbf{2020}, \emph{152}, 071102\relax
\mciteBstWouldAddEndPuncttrue
\mciteSetBstMidEndSepPunct{\mcitedefaultmidpunct}
{\mcitedefaultendpunct}{\mcitedefaultseppunct}\relax
\EndOfBibitem
\bibitem[Parkinson and Oddershede(1997)Parkinson, and
  Oddershede]{Parkinson1997}
Parkinson,~W.~A.; Oddershede,~J. Response function analysis of magnetic optical
  rotation. \emph{International Journal of Quantum Chemistry} \textbf{1997},
  \emph{64}, 599--605\relax
\mciteBstWouldAddEndPuncttrue
\mciteSetBstMidEndSepPunct{\mcitedefaultmidpunct}
{\mcitedefaultendpunct}{\mcitedefaultseppunct}\relax
\EndOfBibitem
\bibitem[Coriani \latin{et~al.}(1997)Coriani, H{\"a}ttig, J{\o}rgensen,
  Halkier, and Rizzo]{Coriani1997}
Coriani,~S.; H{\"a}ttig,~C.; J{\o}rgensen,~P.; Halkier,~A.; Rizzo,~A. {Coupled
  cluster calculations of Verdet constants}. \emph{Chemical Physics Letters}
  \textbf{1997}, \emph{281}, 445--451\relax
\mciteBstWouldAddEndPuncttrue
\mciteSetBstMidEndSepPunct{\mcitedefaultmidpunct}
{\mcitedefaultendpunct}{\mcitedefaultseppunct}\relax
\EndOfBibitem
\end{mcitethebibliography}
